\documentclass[aps,prl,twocolumn,superscriptaddress,amsfont,graphicx,nofootinbib,preprintnumbers]{revtex4-1}%
\UseRawInputEncoding
\usepackage{color,graphicx,epsfig}
\usepackage{ifpdf}
\usepackage{amsmath}
\usepackage{bm}
\usepackage[english]{babel}
\usepackage{amssymb}
\usepackage{braket}
\usepackage{hyperref}
\usepackage{enumerate}
\usepackage{url}

\usepackage{slashed}

\usepackage{changes}

\begin{document}

\title{An Improved Bound on Accelerated Light Dark Matter}

\author{Liangliang Su}
\email{liangliangsu@njnu.edu.cn}
\affiliation{Department of Physics and Institute of Theoretical Physics, Nanjing Normal University, Nanjing, 210023, China}

\author{Lei Wu}
\email{leiwu@njnu.edu.cn}
\affiliation{Department of Physics and Institute of Theoretical Physics, Nanjing Normal University, Nanjing, 210023, China}

\author{Bin Zhu}
\email{zhubin@mail.nankai.edu.cn}
\affiliation{School of Physics, Yantai University, Yantai 264005, China}

\date{\today}

\begin{abstract}
Light (sub-GeV) dark matter has gained increasing interest in terms of direct detection. Accelerated dark matter is a promising candidate that can generate detectable nuclear recoil energy within the sub-GeV range. Because of the large kinetic energy, its interactions with the nucleus are predominantly governed by inelastic scattering, including quasi-elastic and deep inelastic scattering. In this work, we calculated the inelastic effects in dark matter--Earth scattering mediated by a vector particle. Our analysis revealed that the impact of inelastic scattering relies on the mediator mass and the kinetic energy spectrum of dark matter. The results exhibited considerable disparity: the upper bounds of the exclusion limit for the spin-independent cross-section between accelerated dark matter and nuclei via a heavy mediator differ by several tens of times when inelastic scattering is considered.

\end{abstract}
\pacs{Valid PACS appear here}

\maketitle


\section{Introduction}
Multiple cosmological and astrophysical observations provide compelling evidence for the existence of dark matter (DM), which accounts for approximately 26.8\% of the total energy--matter composition of the universe~\cite{ParticleDataGroup:2022pth}. Nevertheless, unraveling the other nature of DM beyond its gravitational interactions remains an unsolved enigma. Consequently, numerous experiments have been devised in the past few decades to detect weakly interacting massive particles (WIMPs)~\cite{Lee:1977ua,Jungman:1995df}, which are considered the most promising candidates for DM. Unfortunately, despite the utilization of these experimental probes, no conclusive signal in support of WIMPs has been found to date~\cite{PandaX-4T:2021bab,LZ:2022ufs,XENON:2023sxq}.

Recently, searching for light DM in (in)direct detection experiments has gained substantial attention~\cite{Essig:2011nj,Hochberg:2015pha,Essig:2015cda,Essig:2017kqs,Knapen:2017xzo,Bertone:2018krk,DAgnolo:2018wcn,PandaX-II:2021lap,Su:2021jvk,Calabrese:2021src,Elor:2021swj,Yang:2022eaq,Calabrese:2022rfa,Ambrosone:2022mvk}. One of the primary challenges associated with the detection of light DM halo is the relatively low nuclear recoil energy generated during DM--target interactions, which rapidly decreases the sensitivity of the detectors employed. Numerous potential experimental setups have been proposed to address this issue, including the use of semiconductors~\cite{Graham:2012su,Hochberg:2016sqx,Bloch:2016sjj,Li:2022fho,Gu:2022vgb,Jia:2022fri} or superconductors~\cite{Hochberg:2016ajh,Hochberg:2019cyy,Griffin:2020lgd,Hochberg:2021yud} featuring lower detection thresholds. Some special processes that do not rely on nuclear recoil signals can also be used to detect light DM halo, such as ``bremsstrahlung'' \cite{Kouvaris:2016afs,McCabe:2017rln,XENON:2019zpr,GrillidiCortona:2020owp} and the ``Migdal'' effect~\cite{Vergados:2005dpd,Moustakidis:2005gx,Ibe:2017yqa,Dolan:2017xbu,Baxter:2019pnz,Essig:2019xkx,Bell:2019egg,Liang:2019nnx,Nakamura:2020kex,Liang:2020ryg,Flambaum:2020xxo,Acevedo:2021kly,Wang:2021oha,Bell:2021ihi,Cox:2022ekg,Berghaus:2022pbu,Li:2022acp,Adams:2022zvg,Bell:2023uvf,Qiao:2023pbw}. Apart from the DM halo, additional sources of DM exist, such as accelerated DM generated via distinct mechanisms. Accelerated DM can provide a considerable amount of recoil energy, enhancing the detection efficiency for light DM. One intriguing acceleration mechanism is the up-scattering of DM halo in the Milky Way by high-energy cosmic rays (CRs), which is known as cosmic-rays up-scatter DM (CRDM)~\cite{Bringmann:2018cvk,Ema:2018bih,Cappiello:2019qsw, Wang:2019jtk, Ge:2020yuf, Xia:2020apm, Ema:2020ulo,Bell:2021xff, Feng:2021hyz, Wang:2021nbf,PandaX-II:2021kai,Maity:2022exk, Nagao:2022azp}. There are also additional interesting mechanisms, including boosted DM (BDM)~\cite{Agashe:2014yua, Berger:2014sqa, Agashe:2015xkj} and atmospheric DM (ADM)~\cite{Alvey:2019zaa, Su:2020zny, Arguelles:2022fqq, Darme:2022bew,Du:2022hms,Sieber:2023nkq}.

Direct detection experiments for DM are typically conducted underground to shield against CRs. However, this shielding also reduces the number of DM particles reaching the detector. This reduction in DM particles is attributed to the Earth-stopping effect~\cite{Kouvaris:2014lpa,Foot:2014osa,Kavanagh:2016pyr, Kavanagh:2017cru}, whereby DM interacts with particles within the Earth as it passes through. Previous studies have primarily estimated the Earth-stopping effect by assuming that only elastic DM--nucleus scattering (ES) occurs during the interaction. However, this may not be applicable to accelerated DM because inelastic DM--nucleus scattering (IES), including quasi-elastic scattering (QES) and deep inelastic scattering (DIS), may become dominant, particularly in the high-kinetic-energy range. This assertion is supported by calculations of neutrino--nucleus scattering. Therefore, when investigating the Earth-stopping effect for accelerated DM, considering the potential contribution of inelastic scattering (IES) is crucial. Recently, in different calculation frameworks, previous studies~\cite{Alvey:2022pad,Su:2022wpj,PandaX:2023tfq} have highlighted the significance of inelastic DM--nucleus scattering in the Earth-stopping effect. 

Su et al.\cite{Su:2022wpj}  presented a general and consistent approach for calculating both QES and DIS using impulse approximation (IA) and the parton model. However, they focused exclusively on ADM interacting with nuclei via a scalar mediator. Building upon previous calculation frameworks~\cite{Su:2022wpj}, we extend our analyses to include more cases of accelerated DM with a vector mediator. In addition, we employed the ``single scatter’’ approximation to model the Earth-stopping effect to simplify the analysis. Through our calculations, we discovered that the DM--nucleus cross-section with a vector mediator exhibits similar behavior to that of a scalar mediator, as observed in a previous work~\cite{Su:2022wpj}. Specifically, the ratio between elastic and inelastic cross-sections is influenced by the mass of the mediator, with IES becoming dominant when the mass of the vector mediator exceeds approximately 1 GeV in a high-kinetic-energy region. For ADM, the constraints on the on-shell mediator mass render the IES of ADM less significant when a vector mediator is involved. Therefore, this study primarily focused on CRDM and BDM. By comparing the expected nuclear recoil signal with data from the Xenon1T experiment, we observed that the importance of inelastic DM-nucleus scattering in the Earth-stopping effect depends not only on the mediator mass but also on the energy spectrum of DM. For instance, if the population of DM predominantly consists of particles with high kinetic energy, such as BDM, the upper bounds of the exclusion limit can be changed by orders of magnitude before and after considering IES. Conversely, for CRDM with a mass below approximately 50 MeV, the inclusion of IES does not lead to a noticeable difference in the upper bounds.

This paper is organized as follows: We introduce two kinds of accelerated DM mechanisms in the second section, including CRDM and BDM. In the third section, we carefully calculated the elastic and inelastic DM--nucleus cross-section using the IA scheme and the parton model. The Earth-stopping model ``single scatter'' approximation is described in the fourth section. Finally, we display the results of the cross-section, the differential flux of DM around the Xenon1T detector, and the exclusion limit of the spin-independent DM--nucleon cross-section in the fifth section. Finally, the concluding remarks are given in the last section.

\section{Accelerated dark matter}\label{sec:2}
The detection of sub-GeV halo DM with low nuclear recoil energy poses challenges for traditional detectors, such as the Xenon1T experiment. Thus, various acceleration mechanisms have been proposed to enhance the nuclear recoil energy of sub-GeV DM within the detector to address the above limitation. In this section, we discuss two of these acceleration mechanisms.

\subsection{Boosted dark matter}

The BDM mechanism involves a two-component DM sector consisting of species $\chi_A$ and $\chi_B$. In this scenario, the heavier DM component $\chi_A$ (with $m_{\chi_A}>m_{\chi_B}$) is assumed to be dominant and does not directly interact with Standard Model (SM) particles. The annihilation process
\begin{equation}
    \chi_{A}\bar{\chi}_A \to \chi_{B}\bar{\chi}_B
\end{equation}
not only ensures its correct thermal relic abundance density but also produces the BDM particles $\chi_{B}$ with the Lorentz factor $\gamma_{m} =m_{\chi_A}/m_{\chi_B}$. Then, these BDMs can be detected by the interactions between them and SM particles. More details in this regard are given in a previous study~\cite{Agashe:2014yua}.

In this framework, the flux of BDM $\chi_{B}$ from the galaxy center (GC) is given by
\begin{equation}
  \frac{\mathrm{d} \Phi_{\mathrm{GC}}}{\mathrm{d} E_{\chi_B}}=\frac{1}{4 \pi}\frac{\left\langle\sigma v\right\rangle}{4m_{\chi_A}^{2}} J(\Omega) \frac{\mathrm{d} N_{\chi_B}}{\mathrm{d} E_{\chi_B}},
\end{equation}
where the factor $1/4$ denotes the Dirac DM, whereas it is $1/2$ for Majorana DM. $E_{\chi_B}$ denotes the total energy of $\chi_B$. $\left\langle\sigma v\right\rangle$ is the thermally averaged cross-section of the annihilation process $\chi_{A}\bar{\chi}_A \to \chi_{B}\bar{\chi}_B$, which is obtained as
\begin{equation}
    \left\langle\sigma v\right\rangle \simeq \frac{1}{8 \pi \Lambda^{4}}\left(m_{\chi_A}+m_{\chi_B}\right)^{2} \sqrt{1-\frac{m_{\chi_B}^{2}}{m_{\chi_A}^{2}}}+\mathcal{O}\left(v^{2}\right),
\end{equation}
where $\Lambda =250$ GeV is assumed to achieve the desired relic density of $\chi_{A}$~\cite{Agashe:2014yua}. The differential energy spectrum of BDM can be written as:
\begin{equation}
\begin{aligned}
\frac{\mathrm{d} N_{\chi_B}}{\mathrm{d} E_{\chi_B}} &= 2 \delta(E_{\chi_B}- m_{\chi_A}) \\
    &\approx  \frac{2}{\sqrt{2\pi} \sigma_{0} m_{\chi_A} } \exp{\left(-\frac{(1-E_{\chi_B}/m_{\chi_A})^2}{2 \sigma_0^2}\right)},
\end{aligned}
\label{eq:BDM_delta}
\end{equation}
with a small standard deviation $\sigma_0$.

The $J(\Omega)$ factor is defined by the DM density $\rho_{\chi}$:
\begin{equation}
    J(\Delta\Omega) = \int_{\Delta\Omega} \mathrm{d} \Omega \int_{\mathrm{l.o.s}} \rho_{\chi}^2(r(x,\theta))\mathrm{d} x,
\end{equation}
where the unit of the $J(\Omega)$ factor is $\mathrm{GeV}^2\mathrm{cm}^{-5}\mathrm{sr}$. $x$ and $r = \sqrt{R_{0}^2+x^2-2xR_{0}\cos \theta}$ are the line-of-sight (l.o.s) distance to the Earth and the galactocentric distance, respectively, where $R_0 = 8.127$ kpc is the distance from the sun to the GC and $\theta$ is the angle between the Earth--GC axis and the l.o.s. For the DM density, we utilized the generalized Navarro--Frenk--White (NFW) profile with the same parameters as those in previous research~\cite{Arguelles:2019ouk}. The three-dimensional integral of the target solid angle was assumed as the All-sky in this work, so the result of the $J$ factor was $2.3\times 10^{23}$  $\mathrm{GeV}^2\mathrm{cm}^{-5}\mathrm{sr}$.

\subsection{Cosmic-rays up-scatter dark matter}
The basic mechanism of CRDM is that the nonrelativistic halo DM particles in the galaxy are accelerated through their scattering with high-energy CRs.

The CR-induced DM flux can be obtained by
\begin{equation}
\begin{aligned}
\frac{\mathrm{d} \Phi_{\chi}}{\mathrm{d}  T_{\chi}} & =\sum_{i} \int \frac{ \mathrm{d}  \Omega }{4 \pi} \int_{\mathrm{l.o.s. }} \mathrm{d} s  \int_{T_{i}^{\min }}^{\infty} \mathrm{d}  T_{i} \frac{\rho_{\chi}}{m_{\chi}} \frac{\mathrm{d} \sigma_{\chi i}}{\mathrm{d}  T_{\chi}} \frac{\mathrm{d}  \Phi_{i}}{\mathrm{d}  T_{i}} \\
& \equiv D_{\text {eff }} \frac{\rho_{\chi}^{\text {local }}}{m_{\chi}} \sum_{i} \int_{T_{i}^{\min }}^{\infty} \mathrm{d}  T_{i} \frac{\mathrm{d} \sigma_{\chi i}}{\mathrm{d} T_{\chi}} \frac{\mathrm{d} \Phi_{i}^{\mathrm{LIS}}}{\mathrm{d}  T_{i}},
\end{aligned}
\end{equation}
where ${\mathrm{d} \sigma_{\chi i}}/{\mathrm{d}  T_{\chi}}$ is the differential cross-section of the scattering between DM and the CR particles $i$ (only proton is considered in this work). In this work, the energy spectrum of CR particles $i$ in the Galactic halo ${\mathrm{d}  \Phi_{i}}/{\mathrm{d}  T_{i}}$ were assumed to be not significantly different from the local interstellar (LIS) population of CRs ${\mathrm{d} \Phi_{i}^{\mathrm{LIS}}}/{\mathrm{d}  T_{i}}$. The latter can be parameterized by their differential intensity, that is, ${\mathrm{d} \Phi_{i}^{\mathrm{LIS}}}/{\mathrm{d}  T_{i}} = 4\pi (\mathrm{d} R_i/\mathrm{d} T_i)(\mathrm{d} I_i/\mathrm{d} R_i) $, where $R_i$ is the rigidity of CR particles~\cite{Boschini:2017fxq}.

In the second line, the effective distance $D_{\text{eff}}$ is defined as
\begin{equation}
    D_{\mathrm{eff}} \equiv \frac{1}{4 \pi \rho_{\chi}^{\text {local }}} \int_{\text {l.o.s }} \rho_{\chi} \mathrm{d} s \mathrm{d} \Omega,
\end{equation}
where $\rho_{\chi}^{\mathrm{local}} = 0.3 \; \mathrm{GeV}/\mathrm{cm}^3$ is the local DM density. Assuming a homogeneous CR distribution and DM density with NFW profile, we take $D_{\mathrm{eff}} = 0.997$ kpc by performing the full l.o.s integration out to 1 kpc.

For the DM with kinetic energy $T_{\chi}$, the minimal incident CR kinetic energy is
\begin{equation}
T_{i}^{\min }=\left(\frac{T_{\chi}}{2}-m_{i}\right)\left[1 \pm \sqrt{1+\frac{2 T_{\chi}}{m_{\chi}} \frac{\left(m_{i}+m_{\chi}\right)^{2}}{\left(2 m_{i}-T_{\chi}\right)^{2}}}\right]  ,
\end{equation}
where $+$ and $-$ signs correspond to the case of $T_{\chi} >2m_{\chi}$ and $T_{\chi} <2m_{\chi}$, respectively. Besides, $T_{i}^{\min }= (m_i+m_\chi)\sqrt{\frac{m_i}{m_\chi}}$ for $T_{\chi} =2m_{\chi}$.
The above discussion pertains to the scenario of elastic DM--proton scattering. Notably, however, the DIS of DM--proton interaction can also produce (semi-)relativistic DM when the protons involved have high energies~\cite{Guo:2020oum}. Further details on this topic are discussed in the subsequent sections.

The kinetic energy of these accelerated DM particles can reach several hundred MeV or even GeV. In this energy range, the interactions between DM and nuclei become more complex, extending beyond elastic scattering. Therefore, in the next section, we analyzed DM--nucleus scattering across different ranges of DM kinetic energy.

\section{DM--nucleus scattering cross-section}\label{sec:3}
In this paper, we assumed that DM interacts with ordinary matter via a vector particle exchange. For example, a dark photon from a spontaneously broken $U(1)^{\prime}$ gauge group in the dark sector interacts with quarks through kinetic mixing~\cite{Fabbrichesi:2020wbt}. Thus, the Lagrangian of relevant interactions is given by
\begin{equation}
    \mathcal{L}_{\mathrm{I}}=A^{\prime}_{\mu}\left(g_{q} \sum_{q} \bar{q} \gamma^{\mu} q+g_{\chi} \bar{\chi} \gamma^{\mu} \chi\right).
\label{eq:L1}
\end{equation}
Here, $g_q = \epsilon Q_q e$ and $g_{\chi}$ are the couplings of dark photon $A^{\prime}_{\mu}$ with quarks and DM, respectively. Herein, $\epsilon$ represents the kinetic mixing parameter, and $Q_q$ corresponds to the charge number of quarks.

\subsection{Coherent elastic DM--nucleus scattering}

For light nonrelativistic DM, such as halo DM, the transfer momentum from DM to the nucleus system is typically small. In this energy region, the dominant contribution of the DM--nucleus cross-section comes from coherent ES. In calculating this process of DM--nucleus interaction, a crucial step is to derive the DM--nucleon interaction from the DM--quarks level interaction; that is,
\begin{equation}
    \mathcal{O}_{q} \stackrel{\text { step (I) }}{\longrightarrow} \mathcal{O}_{N} \stackrel{\text { step (II) }}{\longrightarrow} \mathcal{O}_{\mathcal{A}},
\end{equation}
where $\mathcal{O}_{q,N,\mathcal{A}}$ denote the quark, nucleon($N= n,p$), and nucleus operator, respectively. Step I in the dark photon model, the calculation from the quark level to the nucleon level, has been given previously~\cite{Dent:2015zpa,AristizabalSierra:2018eqm}:
\begin{equation}
\left\langle N\left(p_{f}\right)\left|\bar{q} \gamma^{\mu} q\right| N\left(p_{i}\right)\right\rangle =\mathcal{N}_{q}^{N} \bar{N} \gamma^{\mu} N  ,
\label{eq:step1}
\end{equation}
where $p_{i(f)}$ is the initial(final) state nucleon momentum. $\mathcal{N}_q^N$ can be understood essentially as the number of quarks within the nucleon; that is, $\mathcal{N}_u^p =2$, among others.

Step II includes the following operation:
\begin{equation}
\begin{aligned}
\left\langle \mathcal{A}\left(p^{\prime}\right)\left|\bar{N} \gamma^{\mu} N\right|\mathcal{A}\left(p\right)\right\rangle &=\bar{\mathcal{A}}\left(\gamma^{\mu} F\left(q^{2}\right)+\frac{\sigma^{\mu \nu} q_{\nu}}{2 m_{\mathcal{A}}} F_{1}\left(q^{2}\right)\right) \mathcal{A},
\end{aligned}
\label{eq:step2}
\end{equation}
where the transfer four-momentum  $q = p^{\prime} -p $ is defined by the four momenta of the incoming and outgoing nucleus, $p$ and $p^{\prime}$. In our analysis, we assume that the form factor of the proton and neutron are equal, denoted as $F_N(q^2) = F(q^2)$, and we adopt the Helm form factor. Additionally, the form factor $F_1(q^2)$ accounts for the electric and magnetic form factors that describe the magnetic moment interaction. However, in this particular process, we can neglect the contribution of $F_1(q^2)$ as it is suppressed by $\mathcal{O}(q/m_{\mathcal{A}})$, where $m_{\mathcal{A}}$ is the mass of the nucleus. Therefore, the effective Lagrangian of DM--nucleus interaction via dark photon can be written as
\begin{equation}
\begin{aligned}
\mathcal{L}_{\mathrm{I}}&=A^{\prime}_{\mu}(g_{N} \sum_{N=p,n} \bar{\mathcal{A}} \gamma^{\mu} \mathcal{A} F\left(q^{2}\right)+g_{\chi} \bar{\chi} \gamma^{\mu} \chi) \\
&= A^{\prime}_{\mu}\left(g_{\mathcal{A}}  \bar{\mathcal{A}} \gamma^{\mu} \mathcal{A}  F\left(q^{2}\right)+g_{\chi} \bar{\chi} \gamma^{\mu} \chi\right),
\end{aligned}
\label{eq:L3}
\end{equation}
where $g_\mathcal{A} = Zg_p+(A-Z)g_n$ is the coupling between the dark photon and nucleus, and $g_p=\epsilon e$ and $g_n =0$ are the couplings of the dark photon with proton and neutron, respectively.

Then, in the rest frame of the nucleus, the differential cross-section of the DM--nucleus ES can be written as follows:
\begin{equation}
\begin{aligned}
\frac{\mathrm{d} \sigma_{\mathrm{ES}}}{\mathrm{d} E_R} &=\frac{ g_\mathcal{A}^2g_{\chi}^2 F^2(E_R) }{4\pi (2m_\mathcal{A} E_R+m_{\gamma^{\prime}}^2)^2 (E_{\chi}^2-m_{\chi}^2)} (2m_\mathcal{A} E_{\chi}^2 \\&
+m_\mathcal{A} E_R^2 -2m_\mathcal{A} E_{\chi} E_{R}-m_{\chi}^2E_R-m_\mathcal{A}^2E_R) \\
&= \frac{ A^2   \bar{\sigma}_{\mathrm{n}}m_{\gamma^{\prime}}^4 F^2(E_R) }{4 \mu_{\mathrm{n}}^2 (2m_\mathcal{A} E_R+m_{\gamma^{\prime}}^2)^2 (E_{\chi}^2-m_{\chi}^2)} (2m_\mathcal{A} E_{\chi}^2 \\
&+m_\mathcal{A} E_R^2-2m_\mathcal{A} E_{\chi} E_{R}-m_{\chi}^2E_R-m_\mathcal{A}^2E_R).
\end{aligned}
\label{eq:diff_xsection_es}
\end{equation}
Here, $E_R$, $E_{\chi}$, and $m_{\chi}$ are the nuclear recoil energy, the incoming energy, and the mass of DM, respectively. The momentum-independent DM--nucleon scattering cross-section is defined by the couplings, the mass of the dark photon $m_{\gamma^{\prime}}$, and the reduced mass of DM and nucleon $\mu_{\mathrm{n}}$; that is,
\begin{equation}
    \bar{\sigma}_{\mathrm{n}}= \frac{g_{\mathcal{A}}^2g_{\chi}^2 \mu_{\mathrm{\mathrm{n}}}^2}{A^2\pi m_{\gamma^{\prime}}^4}= \frac{Z^2g_{p}^2g_{\chi}^2 \mu_{\mathrm{\mathrm{n}}}^2}{A^2\pi m_{\gamma^{\prime}}^4} = \frac{Z^2}{A^2} \sigma_p.
    \label{eq:sigma_n}
\end{equation}

In the case of DM with insufficient transfer momentum, the differential cross-section ${\mathrm{d} \sigma_{\mathrm{ES}}}/{\mathrm{d} E_R}$ predominantly contributes to the total DM--nucleus cross-section. However, for light-accelerated DM, particularly those with kinetic energies equal to or more than 1 GeV, the major processes in DM--nucleus scattering are expected to be QES, DIS, and other inelastic processes. This inference is based on comparisons with lepton--nucleus scattering. In the following subsections, we focus on two specific inelastic processes: QES and DIS.

\subsection{Deep inelastic scattering}
\begin{figure}[ht]
  \centering
\includegraphics[width=7.5cm,height=5cm]{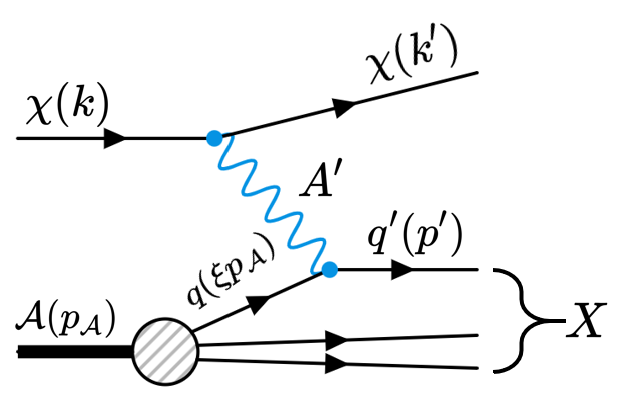}
  \caption{Diagrammatic sketch of DM scatter with the parton in the nucleus.}
  \label{fig:DIS}
\end{figure}

When the incident kinetic energy of DM is extremely high (wherein the mediator wavelength markedly diminishes compared with the nucleon radius), the target nucleus undergoes DIS, resulting in the breakup of the nucleus into a hadronic system $X$. In the calculation of DIS, the parton model is a useful framework. According to this model, the nucleus or nucleon is assumed to consist of quasi-free, point-like parton particles with spin $1/2$, namely, quarks. Within this framework, the DM--nucleus DIS process can be understood as the incoherent superposition of DM--parton elastic scatterings $\chi(k) + q(\xi p_\mathcal{A}) \to \chi(k^{\prime}) +q (p^{\prime})$ (as illustrated in Fig.~\ref{fig:DIS}), where $\xi$ is the momentum fraction of the parton $q$ and $q^{\prime}=q$ in this model.

In the laboratory frame, the spin-average amplitude of DM--parton scattering can be obtained by
\begin{equation}
\begin{aligned}
\overline{|\mathcal{M}(\xi)|^2}&=\frac{g_{\chi}^2g_q^2}{(q^2-m_{\gamma^{\prime}}^2)^2} \\
&\times \mathcal{X}_{\mu \nu} (g^{\mu \alpha} -\frac{q^{\mu}q^{\alpha}}{m_{\gamma^{\prime}}^2}) (g^{\nu \beta} -\frac{q^{\nu} q^{\beta}}{m_{\gamma^{\prime}}^2})  K_{\alpha\beta}.
\end{aligned}
\end{equation}
Here, $\mathcal{X}_{\mu\nu}$ and $K^{\mu\nu}$ represent the current tensor of the DM and parton, respectively, written as
\begin{equation}
\begin{aligned}
\mathcal{X}_{\mu\nu}&=\frac{1}{2}\sum_{spins}[\bar{u}_{\chi}(k^{\prime})\gamma_{\mu}u_{\chi}(k)][\bar{u}_{\chi}(k)\gamma_{\nu}u_{\chi}(k^{\prime})] \\ &=4 k_{\mu}k_{\nu} -2(k_{\mu}q_{\nu}+q_{\mu}k_{\nu})+q^2g_{\mu\nu}; \\
K_{\alpha\beta}&=\frac{1}{2}\sum_{spins}[\bar{u}_q(p^{\prime})\gamma_{\alpha}u_q(\xi p_{\mathcal{A}})][\bar{u}_q(\xi p_{\mathcal{A}})\gamma_{\beta}u_q(p^{\prime})] \\
&=4 \xi p_{{\mathcal{A}}\alpha}\xi p_{{\mathcal{A}}\beta}+2(\xi p_{{\mathcal{A}}\alpha}q_{\beta}+q_{\alpha}\xi p_{{\mathcal{A}}\beta})+q^2g_{\alpha\beta}.
\end{aligned}
\label{eq:DIS_tensor}
\end{equation}
Notably, the electromagnetic current at the dark photon--parton interaction vertex should be conservative; that is, $q^{\alpha} K_{\alpha \beta} =q^{\beta} K_{\alpha \beta} =0$. Although it is obvious here, it is particularly important in the later calculation of QES.

Therefore, the differential cross-section of the DM--nucleus DIS in the parton model can be given by
\begin{equation}
\begin{aligned}
\mathrm{d} \sigma_{\mathrm{DIS}} &= \int_{0}^{1}\frac{\mathrm{d}^3\vec{k}^{\prime}}{(2\pi)^3 2 E_{\chi}^{\prime}}\frac{\mathrm{d}^3\vec{p^{\prime}}}{(2\pi)^3 2 E_{p^{\prime}}}\\
&\times\frac{|\mathcal{M}|^2 (2 \pi)^4 \delta^{4}(\xi p_{\mathcal{A}}+q-p^{\prime})}{4\sqrt{(k\cdot \xi p_{\mathcal{A}})^2-(\xi m_{\mathcal{A}}m_{\chi})^2}} f(\xi)\mathrm{d} \xi \\
&=\frac{\mathrm{d}\omega\mathrm{d}Q^2}{64 \pi m_{\mathcal{A}}^2\omega (E^2_{\chi}-m_{\chi}^2)}\int_{0}^{1}\frac{f(\xi)}{\xi}\mathrm{d} \xi \overline{|\mathcal{M}(\xi)|^2}\delta(\xi-x)\\
&=\sum_q\frac{\mathrm{d}\omega\mathrm{d}Q^2}{32 \pi m_{\mathcal{A}} (E_{\chi}^2-m_{\chi}^2)}\frac{g_{\chi}^2g_q^2}{(Q^2+m_{\gamma^{\prime}}^2)^2} f_{q/A}(x,Q^2) \\
&\times \left[\frac{Q^2(4E_{\chi}(E_{\chi}-\omega)-Q^2)}{\omega^2}+2Q^2-4m_{\chi}^2\right].\\
\end{aligned}
\end{equation}
Here, $q = (\omega, \vec{q})$ and $\omega$ are the transfer momentum and energy, respectively. Meanwhile, the parameter $Q^2$ is defined as $Q^2 \equiv-q^2 = -(k-k^{\prime})^2=2E_{\chi}(E_{\chi}-\omega)-2\vec{k}\vec{k}^{\prime}\cos{\theta}-2m_{\chi}^2$, where $\theta$ is the scattering angle between DM and partons. The function $f_{q/A}(x,Q^2)$, the nuclear parton distribution, describes the probability that the parton carries a fraction $x$ of the nucleus, where $x=Q^2/(2 m_{\mathcal{A}} \omega)$ is called the Bjorken scaling variable~\cite{Buckley:2014ana,AbdulKhalek:2022fyi}.

However, in the rest frame of DM, the differential cross-section is written as
\begin{equation}
\begin{aligned}
    {\mathrm{d}\sigma}_{\mathrm{DIS}}&= \sum_q \frac{\mathrm{d}\omega\mathrm{d}\cos \theta|k^{\prime}|f_{q/A}(x,Q^2) }{8 \pi m_\chi \sqrt{E^2_{\mathcal{A}}-m_{\mathcal{A}}^2}} \frac{g_{\chi}^2g_q^2}{(Q^2+m_{\gamma^{\prime}}^2)^2}\\
    &\times [\frac{8(x m_{\chi} E_\mathcal{A})^2}{Q^2}-4x m_{\chi} E_\mathcal{A}-2m_{\chi}^2-2x^2 m_\mathcal{A}^2+Q^2],
\end{aligned}
\end{equation}
where $Q^2 = 2 m_{\chi} \omega $ and $x = Q^2/(-2p_{\mathcal{A}}\cdot q) >0$.

\subsection{QES in the impulse approximation}
The QES of DM--nucleus also plays an important role in scattering with high transfer momentum, apart from DIS. When the spatial resolution $1/|q|$ is less than the nuclear radius, the DM--nucleus scattering is regarded as the incoherent sum of DM--nucleons scattering rather than the coherent scattering. Herein, the dominant process is assumed to be a nucleon knockout; that is,
\begin{equation}
    \chi (k)+ \mathcal{A} (p_{\mathcal{A}}) \to \chi(k^{\prime}) + X (\to N + Y) .
\label{eq:QES_process}
\end{equation}
Here, the knocked-out nucleon $N=n \; \mathrm{or} \; p$ and the residual nucleus $Y=(\mathcal{A}-1)$ make up the final state $X$ system. The existence of the dark photon--neutron interaction is noteworthy because of the nontrivial magnetic moment distribution of the neutron deviating from a point-like distribution at high transfer momentum. Then, the coupling of the dark photon--neutron is regarded as $g_{n} =\epsilon e =g_{p}$.

Thus, the inclusive differential cross-section of DM--nucleus QES can be written in the Born approximation as
\begin{equation}
\begin{aligned}
   \frac{\mathrm{d} \sigma_{\mathrm{QES}}}{\mathrm{d} E_{\chi}^{\prime} \mathrm{d} \Omega}&= \frac{g_{p}^2 g_{\chi}^2}{16 \pi^2 }\frac{\left|\vec{k}^{\prime}\right|}{|\vec{k}|} D_{\mu \nu}\left(D_{\alpha \beta}\right)^{\dagger} \mathcal{X}^{\mu \alpha} W^{\nu \beta}  \\
   &= \frac{A^2}{Z^2}\frac{\bar{\sigma}_{\mathrm{n}} m_{\gamma^{\prime}}^4}{16 \pi \mu_{\mathrm{n}}^2}\frac{\left|\vec{k}^{\prime}\right|}{|\vec{k}|} D_{\mu \nu}\left(D_{\alpha \beta}\right)^{\dagger} \mathcal{X}^{\mu \alpha} W^{\nu \beta} ,
\end{aligned}
\end{equation}
where the factor $A^2/Z^2$ comes from the definition of $\bar{\sigma}_{\mathrm{n}}$ in Eq~\ref{eq:sigma_n}. The tensor $D_{\mu\nu} =(g_{\mu\nu} -q_{\mu}q_{\nu}/m_{\gamma^{\prime}}^2)$ denotes the propagator of the dark photon. Similarly, we only considered the contribution of the $g_{\mu\nu} g_{\alpha \beta}$ term because the DM current at the DM--dark photon vertex was also conservative ($q_{\mu} \mathcal{X}^{\mu\nu} = q_{\nu} \mathcal{X}^{\mu\nu}=0$), where the DM current tensor $\mathcal{X}^{\mu \nu}$ is given by Eq.~\ref{eq:DIS_tensor}. Besides, the nuclear current tensor $W^{\mu\nu}$ that includes all information on the target nuclear structure is given by
\begin{equation}
\begin{aligned}
   W^{\mu \nu} &=\overline{\sum}_{X}\left\langle \mathcal{A}\left|J^{\mu}_{\mathcal{A}}(0)\right| X\right\rangle\left\langle X\left|J^{\nu}_{\mathcal{A}}(0)\right| \mathcal{A}\right\rangle\\
    &\times \delta^{(4)}\left(p_X+k^{\prime}-p_{\mathcal{A}}-k\right).
\end{aligned}
\end{equation}
Here, $J^{\mu}_{\mathcal{A}}(0)$ is the nuclear current operator.

\begin{figure}[ht]
  \centering
\includegraphics[width=7.5cm, height=5cm]{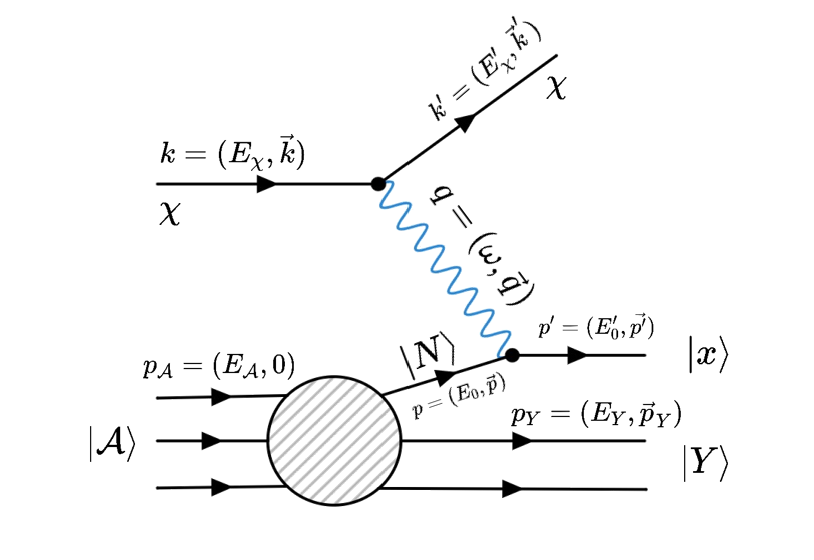}
  \caption{Diagrammatic sketch of DM--nucleus QES.}
  \label{fig:QES}
\end{figure}

In the calculation of the nuclear current tensor $W^{\mu\nu}$, the reliability of nuclear many-body theory, which incorporates the nonrelativistic wave function of the initial and final states, diminishes significantly under conditions of high transfer energy $\omega$ or momentum $q$. Thus, an alternative approach known as the IA scheme~\cite{Benhar:2005dj, Ankowski:2005wi, Ankowski:2007uy, Ankowski:2011ei, Ankowski:2013gha} is employed for the calculation of $W^{\mu\nu}$ and the cross-section of QES in DM--nucleus interactions to address this limitation, particularly in the high-transfer-momentum region ($|\vec{q}| > 350$ MeV). Within the IA scheme, the nucleus is regarded as an assembly of individual nucleons, and DM interacts independently with these nucleons. Meanwhile, the residual nucleus $Y$ is treated as a spectator system (refer to Fig.~\ref{fig:QES}). Additionally, we assumed that the nucleon and residual nucleus evolve autonomously after scattering, disregarding the contributions of final state interactions except for those resulting from Pauli blocking.

Within the IA scheme, the nuclear current operator $J_\mathcal{A}(0)$ can be given by the one-body nucleon current operator~\cite{Benhar:2005dj}
\begin{equation}
    J_{\mathcal{A}}^{\mu}(0) = \sum_{i} j_{i}^{\mu}(0),
\end{equation}
and the final state $|X\rangle$ can be rewritten as
\begin{equation}
|X\rangle \rightarrow|x, \vec{p}^{\;\prime}\rangle \otimes\left|Y, \vec{p}_{Y}\right\rangle.
\end{equation}
where $|x, \vec{p}^{\;\prime}\rangle $ and $\left|Y, \vec{p}_{Y}\right\rangle$ are the knockout nucleon with momentum $\vec{p}^{\;\prime}$ and residual nucleus, respectively. Then, the current matrix elements can be given by
\begin{equation}
\begin{aligned}
\left\langle \mathcal{A}\left|J_{\mathcal{A}}^{\mu}(0)\right| X\right\rangle &\rightarrow\left\langle\mathcal{A}\mid Y, \vec{p}_{Y} ; N,-\vec{p}_{Y}\right\rangle \\
&\times \sum_{i}\left\langle N,-\vec{p}_Y\left|j_{i}^{\mu}(0)\right| x, \vec{p}^{\prime}\right\rangle,
\end{aligned}
\end{equation}
Here, we inserted a complete set of free nucleon states $|N,\vec{p}\rangle$:
\begin{equation}
\begin{aligned}
    \int \mathrm{d}^{3} \vec{p}|N, \vec{p}\rangle\langle\vec{p}, N| =I.
\end{aligned}
\end{equation}
Combining the results, one obtains
\begin{equation}
\begin{aligned}
W^{\mu \nu} &=\sum_{x, Y} \int \mathrm{d}^{3} \vec{p}_{Y} \mathrm{~d}^{3} \vec{p}^{\prime}\left|\left\langle A \mid Y, \vec{p}_{Y} ; N,-\vec{p}_{Y}\right\rangle\right|^{2} \frac{m_{N}^2}{E_0 E_0^{\prime}} \\
& \times \sum_{i}\langle -\vec{p}_{Y}, N|j_{i}^{\mu}(0)| x, \vec{p}^{\prime}\rangle\langle\vec{p}^{\prime}, x|j_{i}^{\nu}(0)| N,-\vec{p}_{Y}\rangle \\
& \times \delta^{3}\left(\vec{q}-\vec{p}_{Y}-\vec{p}^{\prime}\right) \delta\left(\omega+E_{\mathcal{A}}-E_{Y}-E_{0}^{\prime}\right),
\end{aligned}
\end{equation}
where $m_N$ is the mass of the nucleon. ${m_{N}}/{E_0}  $ and ${m_{N}}/{E_0^{\prime}} $ are the implicit covariant normalization factor of the four spinors of the initial and final nucleons.  The  function $\delta\left(\omega+E_{\mathcal{A}}-E_{Y}-E_{0}^{\prime}\right)$ can be rewritten as
\begin{equation}
\begin{aligned}
\delta\left(\omega+E_{\mathcal{A}}-E_{Y}-E_{0}^{\prime}\right) &=\int \mathrm{d} E \delta\left(E-m_{N}+E_{0}-E_{Y}\right) \\
& \times \delta\left(\omega-E+m_{N}-E_{0}^{\prime}\right),
\end{aligned}
\end{equation}
where $E$ is the removal energy that is closely related to the spectral function $P(\vec{p},E)$; that is,
\begin{equation}
\begin{aligned}
P(\vec{p}, E) &=\sum_{Y}|\langle\mathcal{A}\mid Y,-\vec{p}; N, \vec{p}\rangle|^{2} \\
& \times \delta\left(E-m_{N}+E_{0}-E_{Y}\right).
\end{aligned}
\end{equation}
The spectral function contains information on the internal structure of the target nucleus and describes the probability distribution to knock out a nucleon with momentum $\vec{p}$ from the nucleus and leave the residual nucleus $Y$ with energy $E_Y= E_{\mathcal{A}}-m_N+ E$. In this work, the normalization conditions $\int \mathrm{d}^3 \vec{p} \mathrm{d} E P(\vec{p}, E)=1$ must be satisfied.

Then, the nuclear current tensor can be rewritten as
\begin{equation}
\begin{aligned}
  W^{\mu \nu}&=\sum_{i} \int \mathrm{d}^{3} \vec{p} \mathrm{~d} E H_{i}^{\mu \nu} \frac{m_{N}^2}{E_0 E_{0}^{\prime}} P(\vec{p}, E) \\
&\times \delta\left(\omega-E+m_{N}-E_{0}^{\prime}\right),
\end{aligned}
\label{eq:n_tensor}
\end{equation}
where
\begin{equation}
\begin{aligned}
H_{i}^{\mu \nu}(q) &=\sum_{x}\left\langle N,\vec{p}\left|j_{i}^{\mu}\right| x, \vec{p}+\vec{q}\right\rangle\left\langle\vec{p}+\vec{q}, x\left|j_{i}^{\nu}\right| N, \vec{p}\right\rangle \\
& = \frac{1}{2} \mathrm{Tr}\left[\Gamma^{\mu}\frac{\not p+m_N}{2m_N}\gamma^{0} \Gamma^{\nu \dagger}\gamma^{0}\frac{\not p^{\prime}+m_N}{2 m_N}\right] \\
\end{aligned}
\end{equation}
with
\begin{equation}
    \Gamma^{\mu} = \gamma^{\mu}(F_{i,1}(q)+F_{i,2}(q))-\frac{(p+p^{\prime})^{\mu}}{2m_N}F_{i, 2}(q),
\end{equation}
where the form factors $F_{1,i}$ and $F_{2,i}$ are defined as the function of the electric and magnetic nucleon form factors $G_{E}^{N}$ and $G_{M}^N$, respectively (see Ref.~\cite{Benhar:2005dj, Ankowski:2005wi} for more details). Hence, we can obtain the hadronic current tensor
\begin{equation}
\begin{aligned}
        H_{i}^{\mu \nu}(q) &= H_{1,i}\left(-g^{\mu\nu}+\frac{q^{\mu}q^{\nu}}{q^2}\right) \\
        &+\frac{H_{2,i}}{m_N^2}\left(p^{\mu}-\frac{p\cdot q}{q^2}q^{\mu}\right)\left(p^{\nu}-\frac{p\cdot q}{q^2}q^{\nu}\right).
\end{aligned}
\end{equation}
The structure function $H_{1/2,i}$ is given by
\begin{equation}
\begin{aligned}
H_{1,i} &= \tau(F_{1,i}+F_{2,i})^2;\\
H_{2,i} &= F_{1,i}^2+\tau F_{2,i}^2,
\end{aligned}
\end{equation}
where $\tau = -q^2/4m_N^2$.

However, Eq~\ref{eq:n_tensor} is obtained under the assumption of a free nucleon. Thus, $q$ in Eq.~\ref{eq:n_tensor} is required to perform a replacement to patch up the problem of the nuclear binding of the struck nucleon:
\begin{equation}
    q = (\omega,\vec{q}) \rightarrow \tilde{q} \equiv (\tilde{\omega},\vec{q}), \; \mathrm{with}\; \tilde{\omega} =E_{0}^{\prime}-E_{0},
\end{equation}
The hadronic current tensor corresponds to $H^{\mu\nu} (q)\to \tilde{H}^{\mu\nu}(\tilde{q})$. However, this replacement violates the conservation of the electromagnetic current at the dark photon--nucleon interaction vertex; that is, $q_{\mu} \tilde{H}^{\mu\nu}_i \neq 0$. Hence, we must restore it using~\cite{Benhar:2006wy,Ankowski:2007uy}
\begin{equation}
    \mathcal{X}_{\mu\nu} \tilde{H}^{\mu \nu} \rightarrow \mathcal{X}_{\mu\nu} \tilde{H}^{\mu \nu} + \mathcal{X}_{\mu\nu} \tilde{H}^{\mu \nu}_{\mathrm{cor}}
\end{equation}
with
\begin{equation}
    \mathcal{X}_{\mu\nu} \tilde{H}^{\mu \nu}_{\mathrm{cor}} = \frac{c_1}{\tilde{q}^2} \tilde{H}_{1,i} + \frac{c_2}{m_N^2} \tilde{H}_{2,i},
\end{equation}
where the coefficients $c_1$ and $c_2$ are written as
\begin{equation}
\begin{aligned}
c_1 &= (\omega-\tilde{\omega})\left[(\mathcal{B}_1^2-\omega^2)(\omega+\tilde{\omega})-4(|\vec{k}||\vec{q}|\mathcal{B}_1-\omega \vec{k}\cdot\vec{q})\right]\\
c_2 &= c_1 \mathcal{B}_2^{2}+4(\omega-\tilde{\omega}) \mathcal{B}_2\mathcal{B}_1\left[\vec{k}\cdot\vec{p}-\frac{\vec{p}\cdot\vec{q}}{\vec{q}^2}\vec{k}\cdot\vec{q}\right]
\end{aligned}
\end{equation}
with
\begin{equation}
\begin{aligned}
\mathcal{B}_1&= \frac{2\vec{k}\cdot\vec{q}}{|\vec{q}|}-|\vec{q}|;\\
\mathcal{B}_2&= \frac{1}{2|\vec{q}|} (2 E_{0}+\omega).
\end{aligned}
\end{equation}

To sum up, the inclusive differential cross-section of DM--nucleus QES can rewritten as
\begin{equation}
\begin{aligned}
 \frac{\mathrm{d} \sigma_{\mathrm{QES}}}{\mathrm{d} E_{\chi}^{\prime} \mathrm{d} \Omega}  & = Z\frac{\mathrm{d} \sigma_p}{\mathrm{d} E_{\chi}^{\prime} \mathrm{d} \Omega}+(A-Z)\frac{\mathrm{d} \sigma_n}{\mathrm{d} E_{\chi}^{\prime} \mathrm{d} \Omega}
\end{aligned}
\end{equation}
with
\begin{equation}
\begin{aligned}
    \frac{\mathrm{d} \sigma_N}{\mathrm{d} E_{\chi}^{\prime} \mathrm{d}  \Omega} &= \frac{A^2}{Z^2}\frac{\bar{\sigma}_{\mathrm{n}} m_{\gamma^{\prime}}^4}{16 \pi \mu_{\mathrm{n}}^2  (q^2-m_{\gamma^{\prime}}^2)^2}\frac{\left|\vec{k}^{\prime}\right|}{|\vec{k}|} \int \mathrm{d}^{3} \vec{p} \mathrm{~d} E \frac{m_{N}^2}{E_0 E_{0}^{\prime}} \\
    &\times P(\vec{p}, E) \delta\left(\omega-E+m_{N}-E_{0}^{\prime}\right)  \mathcal{X}_{\mu \nu} \tilde{H}^{\mu\nu}_N.
\end{aligned}
\end{equation}

\section{Earth Stopping}\label{sec:4}
\begin{figure*}[t]
  \centering
\includegraphics[width=8.3cm,height=6.1cm]{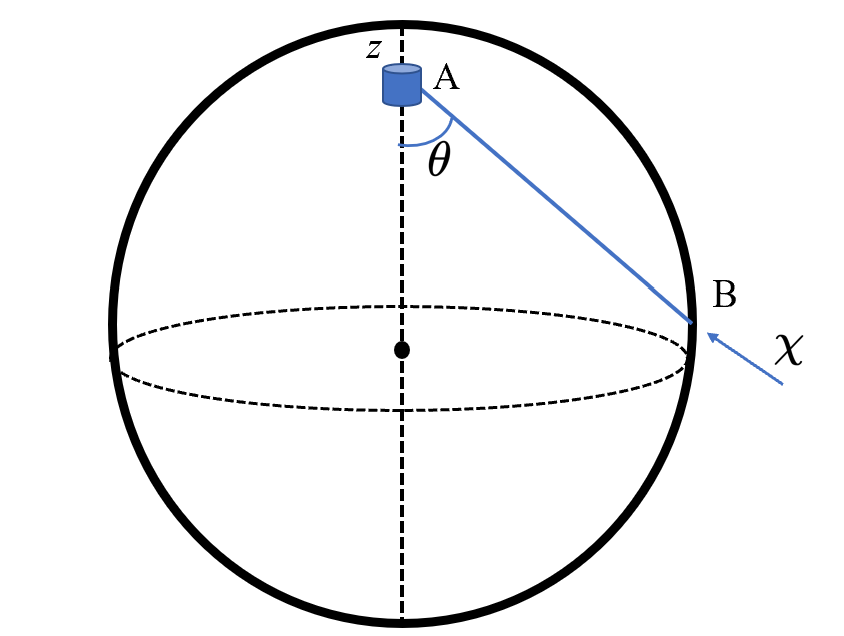}
\includegraphics[width=8cm,height=6.3cm]{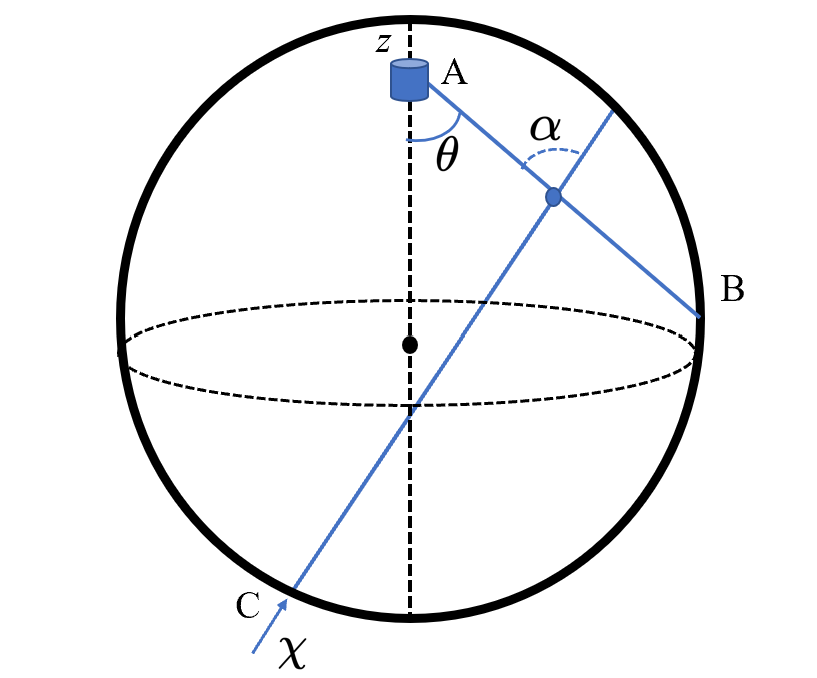}
  \caption{Geometric schematic diagram for attenuation (left plane) and deflection (right plane) as DM passes through the Earth.}
  \label{fig:Earth-stopping}
\end{figure*}
In this work, we only considered the contribution of attenuation because it is dominant for the spin-independent sub-GeV DM in ``single-scatter'' approximation ~\cite{Kavanagh:2016pyr}. The DM not only interacts with nuclei in the detector but also with the Earth's media as they pass through the Earth. Therefore, some DM particles fail to reach the detector or lose part of their energy in the so-called Earth-stopping effect. In this work, we considered a simplified approximate model, the ``single-scatter'' approximation, which assumes that DM particles scatter with the Earth at most once. In this model, the differential flux of DM around the detector located at a depth $z$ contains two contributions:
\begin{equation}
    \frac{\mathrm{d} \Phi_{\chi}^z}{\mathrm{d} T_{\chi} } = \left.\frac{\mathrm{d} \Phi_{\chi}^z}{\mathrm{d} T_{\chi} }\right|_{\mathrm{A}}  + \left.\frac{\mathrm{d} \Phi_{\chi}^z}{\mathrm{d} T_{\chi} }\right|_{\mathrm{D}} ,
\end{equation}
Here, the first and second terms imply the attenuated and deflected population of DM particles, respectively, as shown in figure.~\ref{fig:Earth-stopping}. In this work, we only considered the attenuation contribution $\mathrm{d}\Phi^z_{\chi}/\mathrm{d}T_{\chi}|_\mathrm{A}$ in Earth-stopping because of the negligible contribution of deflection for spin-independent DM~\cite{Kavanagh:2016pyr}.

The differential flux of DM around the detector from the attenuation effect is given by
\begin{equation}
\begin{aligned}
\frac{\mathrm{d} \Phi_{\chi}^{z}}{\mathrm{d} T_{\chi}^{z}} &= \int  \mathcal{P}_{\mathrm{surv}}(T_{\chi}, \cos \theta)  \frac{\mathrm{d} \Phi_{\chi}}{\mathrm{d} T_{\chi} \mathrm{d} \Omega } \mathrm{d} \Omega\\
&= 2 \pi \int_{-1}^{1}  \mathcal{P}_{\mathrm{surv}}(T_{\chi}, \cos \theta)  \frac{\mathrm{d} \Phi_{\chi}}{\mathrm{d} T_{\chi}\mathrm{d} \Omega } \mathrm{d} \cos \theta,
\end{aligned}
\label{eq:single_flux}
\end{equation}
where $T_{\chi}^z$ is the kinetic energy of the DM particles reaching the detector.

\begin{table}[ht]
\centering
\begin{tabular}{lccc}
\hline & & \multicolumn{2}{c}{\text{Mass\;\;fraction}} \\
\cline{3-4}
\text { Element } & $\bar{n} \left(\text{cm}^{-3}\right)$ &  \text { Core } & \text { Mantle }  \\
\hline \text { Iron }  & $6.11 \times 10^{22}$ & 0.855 & 0.0626 \\
\text { Oxygen }  & $3.45 \times 10^{22}$ & 0.0 & 0.440 \\
\text { Silicon }  & $1.77 \times 10^{22}$ & 0.06 & 0.210 \\
\text { Magnesium } & $1.17 \times 10^{22}$ & 0.0 & 0.228 \\
\text { Sulfur }  & $2.33 \times 10^{21}$ & 0.019 & 0.00025 \\
\text { Aluminum }  & $1.09 \times 10^{21}$ & 0.0 & 0.0235 \\
\text { Calcium }  & $7.94 \times 10^{20}$ & 0.0 & 0.0253 \\
\text { Sodium }  & $1.47 \times 10^{20}$ & 0.0 & 0.0027 \\
\hline \hline
\end{tabular}
\caption{Composition of the Earth’s core and mantle~\cite{MCDONOUGH2003547}. The second column displays the average number density $\bar{n}$ of the constituent elements, whereas the third and fourth columns delineate the mass fractions of these elements in the Earth's core and mantle, respectively.}
\label{tab:Earth_element}
\end{table}

During the passage of DM particles through the Earth, the survival probability $\mathcal{P}_{\mathrm{surv}}$ for the DM with kinetic energy $T_{\chi}$ is defined as
\begin{equation}
\begin{aligned}
\mathcal{P}_{\mathrm{surv}}(T_{\chi}, \cos \theta) &=\exp \left[-\int_{\mathrm{BA}} \frac{\mathrm{d} l}{\lambda(\mathbf{r}, T_{\chi})}\right]\\
    &=\exp \left[-\sum_{i} \sigma_{i}^{tot}(T_\chi) \int_{\mathrm{BA}} n_{i}(\mathbf{r}) \mathrm{d} l\right]\\
    &=\mathrm{exp} \left(-\sum_{i} \frac{d_{\mathrm{eff},i}(\cos \theta)}{\bar{\lambda}_i(T_{\chi})} \right),
\end{aligned}
\label{eq:sur_p}
\end{equation}
where the number densities of Earth species $i$ and $n_i$ are assumed to be isotropic. In the last line, we defined the average mean free path $\bar{\lambda}_i = [\sigma_i^{tot} (T_{\chi})\bar{n}_i]^{-1}$ with the average number density
\begin{equation}
    \bar{n}_i \equiv \frac{1}{R_E} \int_{0}^{R_E} n_{i}(r) \mathrm{d}r ,
\end{equation}
where $R_E = 6378.14 \; \mathrm{km}$ is the Earth's radius.
Table~\ref{tab:Earth_element} shows the composition of the Earth. In this table, oxygen and iron are the most abundant elements. The number density profiles of these elements were obtained from previous studies~\cite{MCDONOUGH2003547,Lundberg:2004dn}.

The effective Earth-crossing distance, $d_{\mathrm{eff},i}(\cos \theta)$, is defined by
\begin{equation}
\begin{aligned}
d_{\mathrm{eff},i} & = \frac{1}{\bar{n}_i} \int_{\mathrm{BA}} n_{i} (r) \mathrm{d} l \\
&\approx  \left\{\begin{matrix}
 \displaystyle\int_{R_E \sin\theta}^{R_E}  \frac{2r n_i(r) \mathrm{d} r}{\bar{n}_i\sqrt{r^2-R_E^2\sin^2\theta}} ;& \theta \in \left[0,\dfrac{\pi}{2}\right]\\
 \displaystyle\int_{R_E-z_D}^{R_E} \frac{n_{i}(r)}{\bar{n}_i} \mathrm{d} r ,& \theta \in \left[\dfrac{\pi}{2},\pi\right]
\end{matrix}\right.
\end{aligned}
\end{equation}
Here, the length of line BA, $l$, is the function of $\theta$, which is the angle between the Earth's core-detector direction and the trajectory of DM $\mathrm{B\to A}$, as shown in the left plane in Fig.~\ref{fig:Earth-stopping}.

\section{Results}\label{sec:5}
\begin{figure*}[t]
  \centering
\includegraphics[height=12cm,width=12cm]{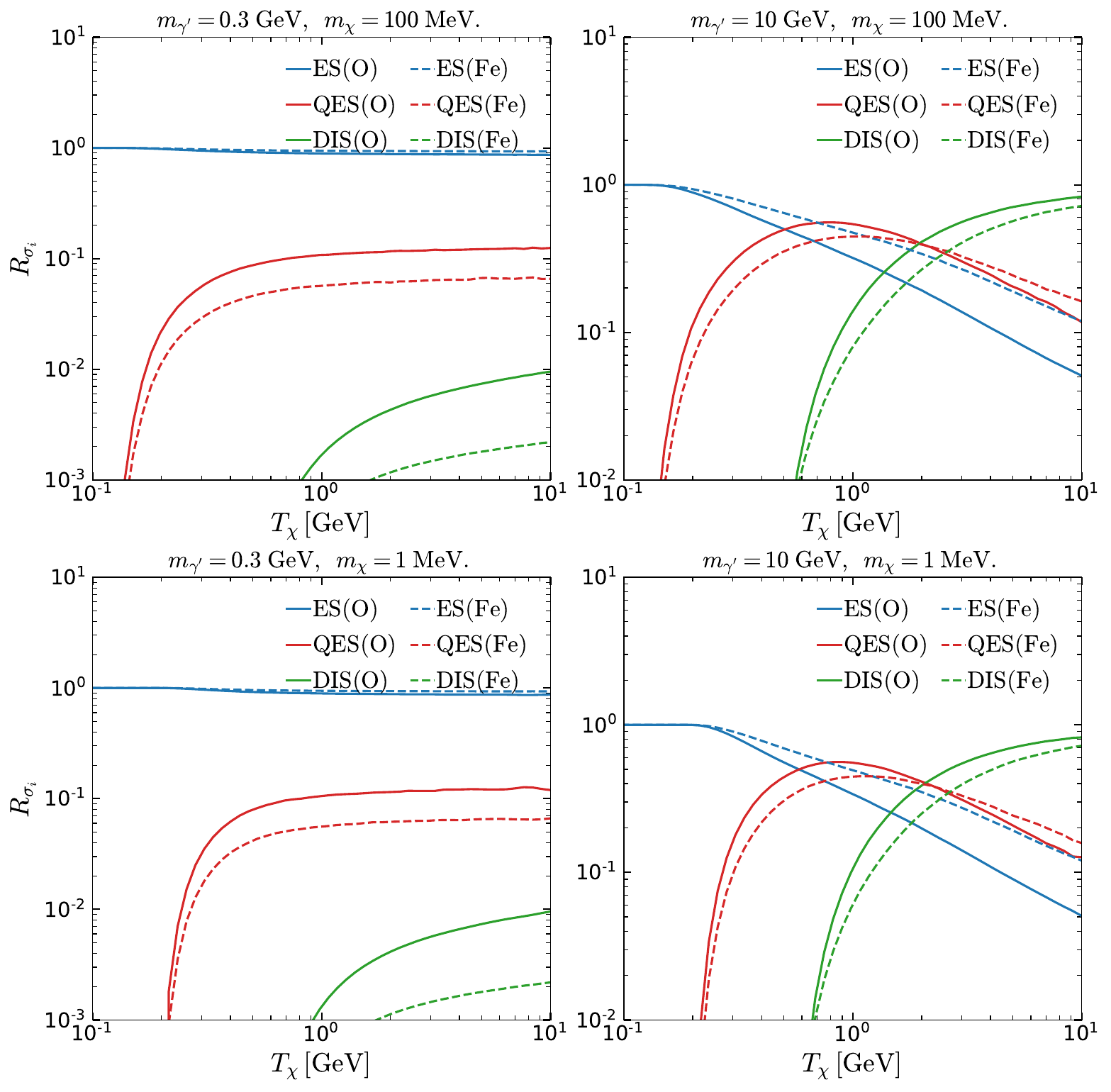}
\caption{Trend of the scattering cross-section ratio $R_{\sigma_i}$ as a function of DM kinetic energy $T_{\chi}$ for different DM and medium masses. Solid and dashed lines represent the results of the interactions of DM with oxygen and iron, respectively.}
\label{fig:sigma_ratio}
\end{figure*}

\subsection{Cross-section}
The ratio of cross-sections is defined to provide a clearer understanding of the relative importance of ES and IES processes in DM--nucleus interactions:
\begin{equation}
    R_{\sigma_i} =\frac{\sigma_i}{\sigma_{tot}},
\label{eq:ratio}
\end{equation}
where $i=\mathrm{ES},\mathrm{QES},\mathrm{DIS}$, and $\sigma_{tot}$ is the sum of the cross-sections of these three processes.

The ratio of cross-sections serves as a valuable indicator for identifying the dominant scattering process between DM particles and nuclei in different kinetic energy regimes. Figure~\ref{fig:sigma_ratio} presents the ratios $R_{\sigma_i}$ for DM scattering with oxygen nuclei (solid lines) and iron nuclei (dashed lines) considering various combinations of DM and mediator masses. For instance, the upper left panel shows $m_{\gamma^{\prime}} = 0.3 $ GeV and $m_{\chi}=100$ MeV as representative parameters for the analysis. The analysis reveals that the dominant process of DM--nucleus scattering is primarily determined by the mediator mass. Specifically, for a mediator mass of $0.3$ GeV, the ratio of QES (red lines) and DIS (green lines) is considerably lower than that of ES in the DM kinetic energy range of interest. However, as the mediator mass increases, IES processes (QES or DIS) become dominant in specific regions of DM kinetic energy. For instance, when considering a mediator with $m_{\gamma^{\prime}} = 10$ GeV, the cross-section of QES exhibits the highest value within the range of several hundred MeV to approximately 2 GeV of DM kinetic energies. This effect can be attributed to the behavior of the propagator, $\mathrm{d}\sigma_i \propto 1/(Q^2+m_{\gamma^{\prime}}^2)^2$, which implies that the differential cross-section of ES is enhanced by a factor of $1/Q_{\mathrm{ES}}^4$ when the mediator's mass is significantly smaller than the typical transfer momentum $Q_{\mathrm{ES}}$. This principle applies similarly to DIS.

Additionally, the DM mass plays a crucial role in determining the threshold for the inelastic process. Specifically, when $m_{\gamma^{\prime}}=0.3$ GeV and $m_{\chi} = 100 $ MeV, the ratio of the QES cross-section can reach up to $0.1\%$ at a kinetic energy of around 130 MeV. Conversely, if DM has $m_{\chi}=1$ MeV, it requires a kinetic energy of over 200 MeV to attain the same ratio. This phenomenon arises from the fact that heavier DM carries more momentum for the same kinetic energy, as manifested by the equation $\vec{p}_{\chi}^2= T_{\chi}^2 +2T_{\chi}m_{\chi}$, making QES and DIS more probable.

In conclusion, the dominant process of DM--nucleus scattering is determined by the specific properties of the DM particle and its mediator. However, the method utilized previously~\cite{Alvey:2022pad}, which involved rescaling the neutral current neutrino--nucleus cross-section to calculate the DM--nucleus scattering, did not lead to this conclusion because of the specific masses of neutrinos and the $Z$ boson used in that analysis.

\subsection{Differential flux}

As shown in Figure.~\ref{fig:sigma_ratio}, inelastic DM--nucleus scattering is dominant for high-kinetic-energy DM with heavy mediator mass. Thus, in this section, we demonstrate the impact of the DM--nucleus IES on the direct detection of CRDM and BDM.

\begin{figure}[ht]
  \centering
\includegraphics[width=8cm,height=5.7cm]{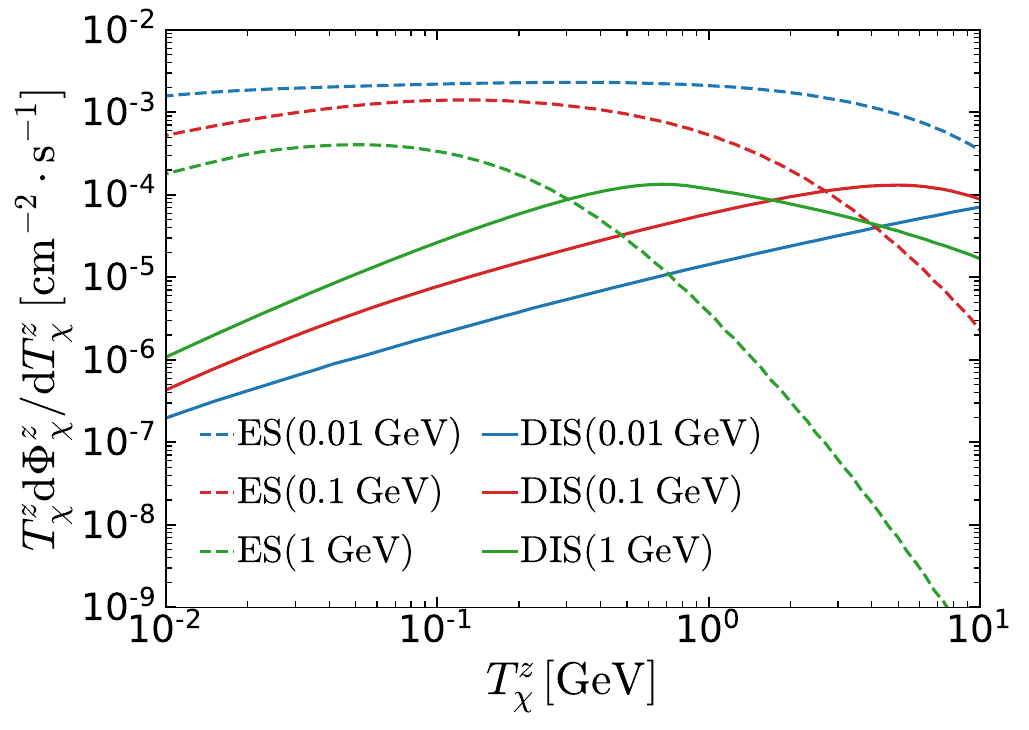}
  \caption{Flux of CRDM on the Earth's surface from ES (dotted lines) and DIS (solid lines) for different DM masses: 0.01 GeV (blue lines), 0.1 GeV (red lines), and 1 GeV (green lines).}
  \label{fig:flux_CRDM_0}
\end{figure}

First, for the CRDM produced by high-energy CRs colliding with DM, the DIS also contributes significantly to the production of abundant DM particles with high kinetic energy, as depicted in Figure~\ref{fig:flux_CRDM_0}. In this process, the flux of DM from DIS is suppressed in the low $Q^2 = 2 m_{\chi} T_{\chi} $ region due to the parton distribution function $f_{q/p}(x,Q^2)$. Consequently, the contribution of DM is relatively poor for DM with lower mass or kinetic energy, as exemplified by the 0.01 GeV DM case (blue solid line). Therefore, this contribution is considered in this work.

\begin{figure}[ht]
  \centering
\includegraphics[width=8cm,height=5.7cm]{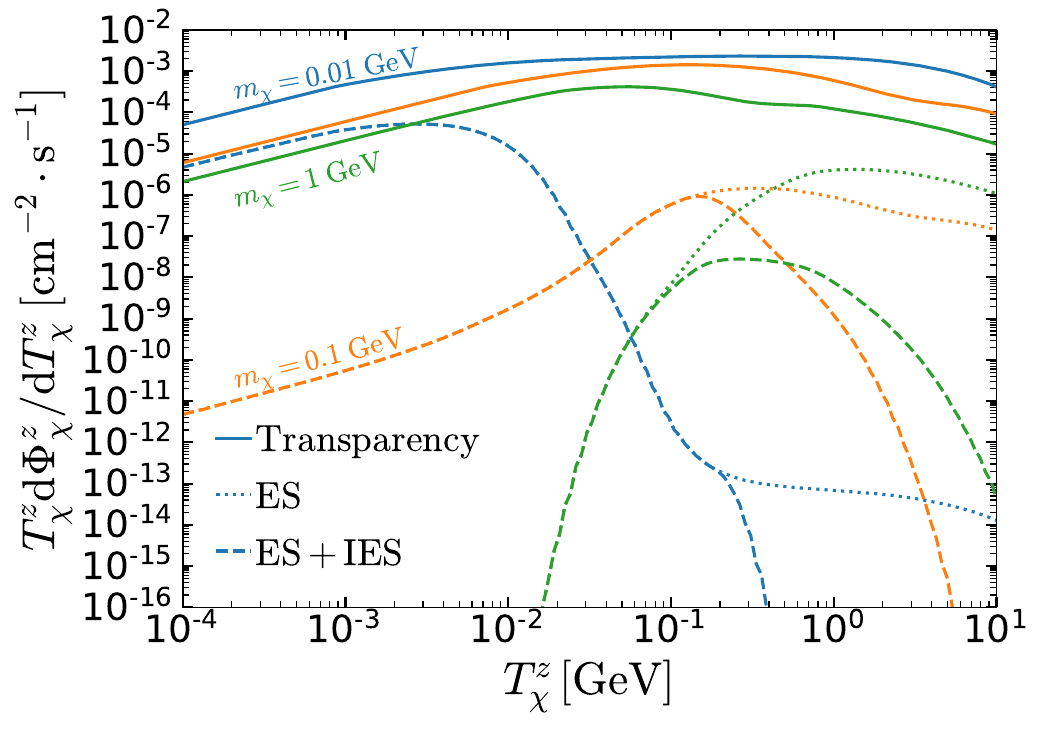}
  \caption{Differential flux of CRDM around the Xenon1T detector for different DM masses $m_{\chi} =0.01$ GeV (blue lines), 0.1 GeV (orange lines), and 1 GeV (green lines) without (dashed lines) and with IES (dotted lines). The corresponding cross-sections are $\sigma_p= 2\times 10^{-28} \; \mathrm{cm}^2$, $2\times 10^{-27} \; \mathrm{cm}^2$, and $2\times 10^{-26}\; \mathrm{cm}^2$, respectively. Here, we set the mediator mass to $m_{\gamma^{\prime}} =10 $ GeV. Besides, the solid lines denote the differential flux of CRDM of the transparent Earth case.}
  \label{fig:flux_CRDM}
\end{figure}

\begin{figure}[ht]
  \centering
\includegraphics[width=8cm,height=5.7cm]{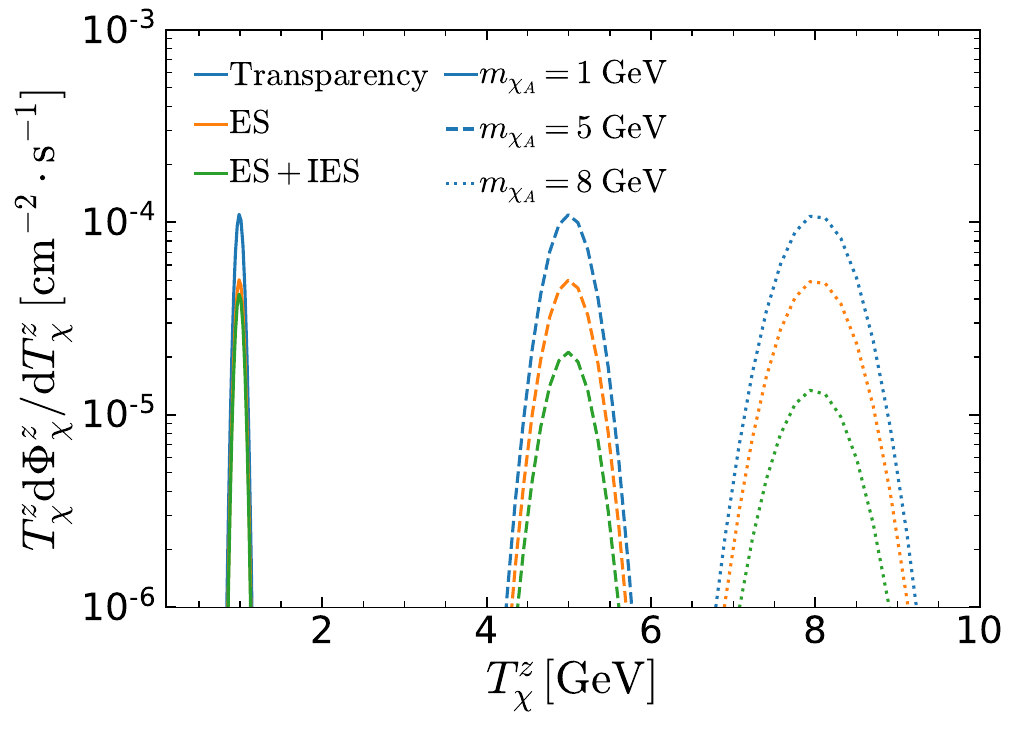}
  \caption{Differential flux of BDM $\chi_B$ around the Earth's surface (blue lines) and the Xenon1T detector with (green lines) and without (orange lines) IES. We take the standard deviation $\sigma_0=0.05$ in Eq.~\ref{eq:BDM_delta}, $m_{\gamma^{\prime}}= 10$ GeV, $\bar{\sigma}_{\mathrm{n}} = 10^{-31}\;\mathrm{cm}^2$, and $m_{\chi_B}=10$ MeV with three heavy DM masses $m_{\chi_A} =1$ GeV (solid lines), 5 GeV (dashed lines) and 8 GeV (dotted lines) as the benchmark points.}
  \label{fig:flux_BDM}
\end{figure}

However, in Earth-stopping processes, CRDM and BDM both may inelastically scatter with particles in Earth. Thus, Figs.~\ref{fig:flux_CRDM} and ~\ref{fig:flux_BDM} respectively display the expected differential fluxes of CRDM and BDM at the Xenon1T detector both before and after incorporating the IES of DM--nucleus interactions with a heavy vector mediator $m_{\gamma^{\prime}} =10$ GeV.

Fig.~\ref{fig:flux_CRDM} demonstrates that the flux of high-kinetic-energy CRDM particles (above approximately $\mathcal{O}(100)$ MeV) for all sub-GeV DM is further suppressed when considering the DM--nucleus IES. This observation aligns with the findings presented in Fig.~\ref{fig:sigma_ratio}, where the flux of DM is exponentially reduced by the total cross-section under the ``single scatter'' approximation, as described in Eqs.~\ref{eq:single_flux} and \ref{eq:sur_p}. In addition, for DM with low mass and high kinetic energy, there is an enhancement in the cross-section of DM--nucleus ES due to the momentum transfer effect during the scattering process (see~\cite{Flambaum:2020xxo}). Consequently, the flux of the 0.01 GeV DM particles is significantly smaller than the fluxes of the other two mass values. In the low-kinetic-energy range, there are also significant differences in the fluxes of different CRDM masses after reaching the detector. For instance, a 1 GeV CRDM particle with kinetic energy below 10 MeV is unlikely to reach the detector. However, approximately one-tenth of a 10 MeV CRDM particle can pass through the Earth with the same coupling parameter. According to Eq.~\ref{eq:diff_xsection_es}, in this kinetic energy region, the cross-section of DM--nucleus ES (the dominant process) increases with the DM mass because a heavier DM particle can transfer more energy during the scattering process; that is, $\sigma_{\mathrm{ES}} = \int^{E_R^{\max}}_0 \frac{\mathrm{d}\sigma_{\mathrm{ES}}}{\mathrm{d} E_{R}} \propto E_{R}^{\max} \frac{\mathrm{d}\sigma_{\mathrm{ES}}}{\mathrm{d} E_{R}}|_{E_R=0}$. Therefore, under the ``single-scatter’’ approximation, the total flux of heavier CRDM is more significantly affected by Earth's stopping when the IES of DM--nucleus interactions is considered.

The conclusion regarding BDM differs from that of CRDM because of its unique flux characteristics, specifically its relatively narrow energy spectrum with a standard deviation of $\sigma_0 = 0.05$, as described in Eq.~\ref{eq:BDM_delta}. The energy distribution of BDM is concentrated around the mass of the heavier DM particle $m_{\chi_A}$. Consequently, when $m_{\chi_A}$ is set to 1 GeV or higher, the dominant process becomes IES rather than ES for BDM--nucleus interactions with a heavy vector mediator $m_{\gamma^{\prime}} = 10$ GeV. Moreover, the ratio of cross-sections between inelastic and elastic scattering gradually increases with the kinetic energy of BDM within the energy range of interest. This behavior is illustrated in Fig.~\ref{fig:sigma_ratio}. Consequently, the survival probability of BDM with larger $m_{\chi_A}$ is smaller, leading to a more pronounced difference in the flux before and after considering the IES, which is given in Fig.~\ref{fig:flux_BDM}.

In conclusion, the impact of inelastic DM--nucleus scattering depends on the energy spectrum of the DM flux. If a significant proportion of DM has high kinetic energy, such as in the case of BDM, the effect is more substantial. Conversely, if the proportion of high-kinetic-energy DM is lower, as seen in CRDM, the impact is relatively smaller.

\subsection{Exclusion limit}

\begin{figure*}[ht]
  \centering
\includegraphics[height=5.6cm,width=17cm]{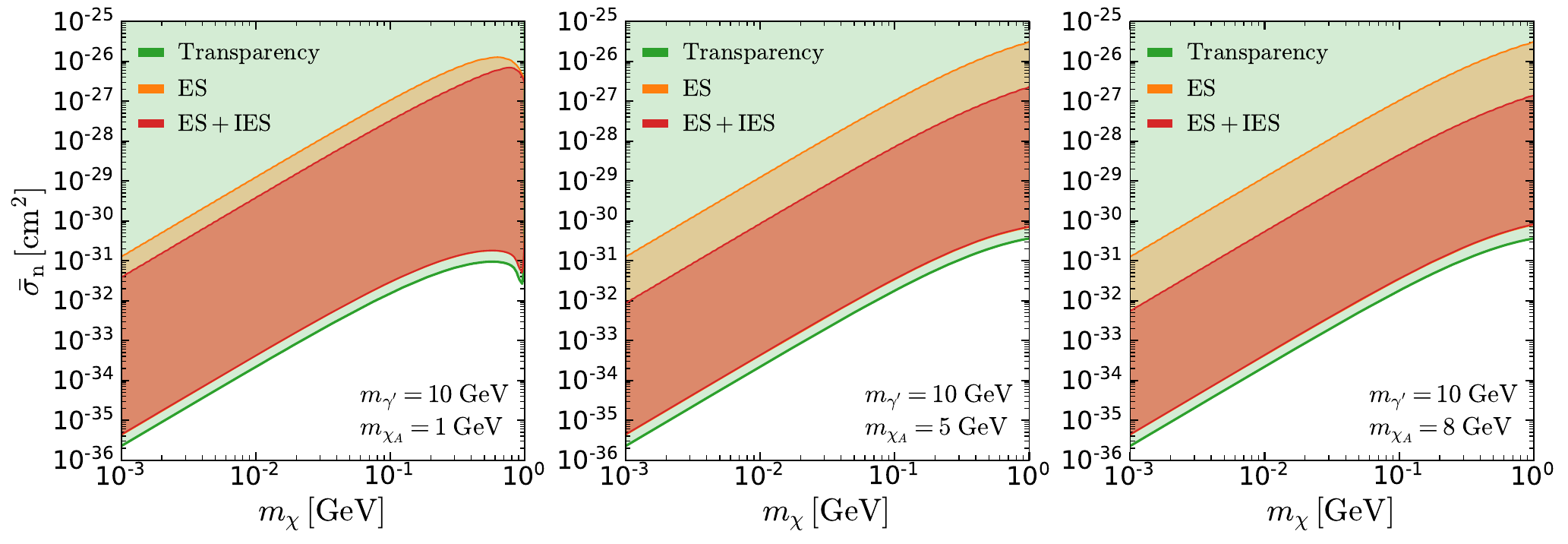}
\caption{Exclusion limit of BDM--nucleon spin-independent cross-section $\bar{\sigma}_{\mathrm{n}}$ for different scattering types: transparent Earth (green) and without (orange) and with (red) IES in Earth stopping. Here, we take the vector mediator mass as 10 GeV and three heavier DM masses $m_{\chi_A} =1$ GeV (left plane), 5 GeV (middle plane), and 8 GeV (right plane).}
\label{fig:limit_BDM}
\end{figure*}

\begin{figure}[ht]
  \centering
\includegraphics[width=7.5cm,height=7.5cm]{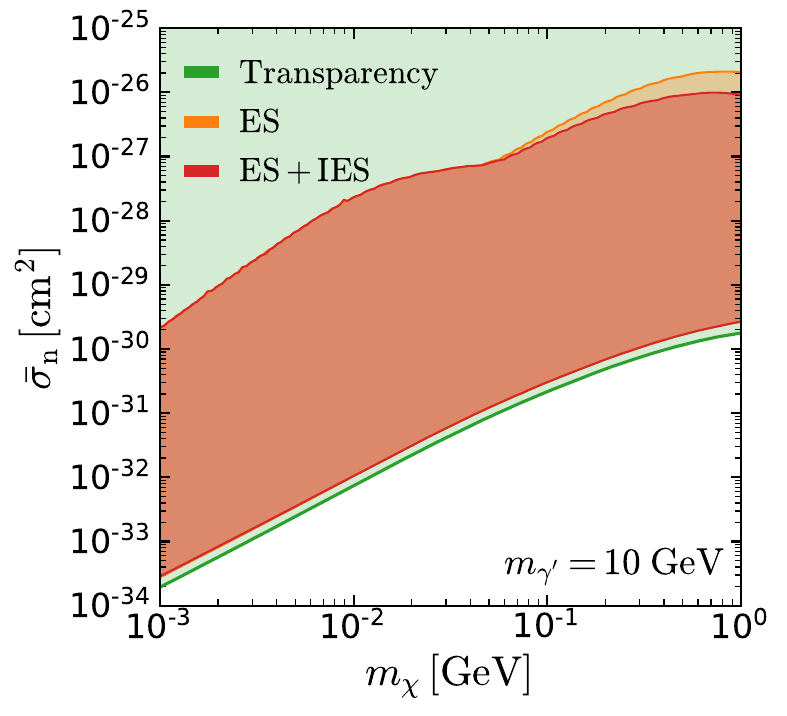}
  \caption{Exclusion limit of CRDM--nucleon spin-independent cross-section $\bar{\sigma}_{\mathrm{n}}$ for different scattering types: transparent Earth (green) and without (orange) and with (red) IES in Earth stopping. Here, we take the vector mediator mass as 10 GeV.}
  \label{fig:limit_CRDM}
\end{figure}

Having obtained the differential flux of DM around the detector, we can proceed to the estimation of the nuclear recoil rate resulting from the interaction between DM and the target:
\begin{equation}
\mathcal{R} = N_T \int_{E_R^{\min}}^{E_R^{\max}}  \mathrm{d} E_R \int_{T_{\chi}^{z,\min}}^{T_{\chi}^{z,\max}} \epsilon(E_R) \frac{\mathrm{d} \Phi_{\chi}^{z}}{\mathrm{d} T_{\chi}^{z}} \frac{\mathrm{d} \sigma}{\mathrm{d} E_R} \mathrm{d} T_{\chi}^{z},
\end{equation}
Here, the number density of the target $N_T$ and the detector's efficiency $\epsilon(E_R)$ are experimentally dependent. In this work, the focus is on the nuclear recoil signal within a specific energy range: $E_R$ ranging from 4.9 keV to 40.9 keV in the Xenon1T experiment. During this process, only the contribution of elastic DM--xenon scattering is considered. This is because the transfer momentum associated with $E_R^{\max}$ is approximately 100 MeV, and IES has a small contribution for transfer momenta below about 200 MeV. In other words, the transfer momentum corresponding to this particular recoil energy region is unlikely to trigger significant IES events. Hence, the analysis concentrated solely on the elastic scattering contribution. Then, the exclusion limits of BDM and CRDM--nucleon spin-independent cross-section $\bar{\sigma}_{\mathrm{n}}$ are shown in Figs.~\ref{fig:limit_BDM} and~\ref{fig:limit_CRDM}, respectively.

Fig.~\ref{fig:limit_BDM} shows that IES contributes equally to all BDM masses at a concrete $m_{\chi_A}$, except for $m_{\chi_B} \approx m_{\chi_A}$. When the mass $m_{\chi_B}$ approaches the mass of $\chi_A$, the exclusion limits tend to tighten because of the amplification of the differential cross-section $\mathrm{d}\sigma \sim \frac{E_{\chi}^2}{E_{\chi}^2 - m_{\chi}^2} = \frac{1}{v_{\chi}^2}$. However, the energy spectrum of BDM depends on $m_{\chi_A}$ and decides the significance of IES, leading to the contribution of IES in Earth-stopping becoming different for various $m_{\chi_A}$. This is why the larger $m_{\chi_A}$ resulted in a wider gap between the upper bounds before and after considering IES.

However, for lighter CRDM, most of the flux in the high-kinetic-energy region is suppressed by elastic scattering due to the momentum transfer effect. Therefore, the recoil signal almost originates from CRDM with kinetic energy from $T_{\chi}^{z,\min} \sim \mathcal{O}(0.01)$ GeV to 0.1 GeV, where IES hardly ever happens. Thus, inelastic DM--nucleon scattering in Earth stopping has little impact on the exclusion limit of CRDM with a mass below about 50 MeV. However, with increasing CRDM mass, the momentum transfer effect gradually weakens, causing the contribution of IES in $T_{\chi}^{z} \in [0.1, 1]$ GeV to emerge. The significance of IES gradually enhances with increasing mass in $m_{\chi}\gtrsim 50$ MeV, and the upper bound of the exclusion limit of 1 GeV CRDM has twice the difference between before and after including IES. This conclusion is consistent with that in Fig.~\ref{fig:flux_CRDM}. Meanwhile, for CRDM with kinetic energies $T_{\chi} > 1$ GeV resulting from the DIS between DM and CRs, there is almost no contribution to the recoil event in the elastic scattering region $E_R \in [4.9,40.9]$ keV. If the experiment focuses on higher energy deposition, this contribution becomes important.

In the case of ``single-scatter’’ approximation, there is typically a factor of two difference between the lower bounds for (in)elastic scattering and transparency. This difference stems from the fact that DM particles are more likely to be scattered away as they traverse through the hemisphere behind the detector. Consequently, under the ``single-scatter’’ approximation, approximately half of the DM particles cannot reach the detector, whereas the transparency approximation assumes that all particles reach the detector without any scattering. Besides, the upper limits depicted in Figs.~\ref{fig:limit_BDM} and \ref{fig:limit_CRDM} can attain approximately $\mathcal{O}(10^{-26})\; \mathrm{cm}^2$. Notably, these large DM interaction cross-sections are subject to constraints imposed by various cosmological and astronomical observations, including Big Bang nucleosynthesis and cosmic microwave background \cite{Gluscevic:2017ywp,Xu:2018efh,Ooba:2019erm}. While relativistic CRDM and BDM may have been produced in the early universe, their influence on large-scale structures can be disregarded because of their comparatively negligible population when contrasted with cold DM.

\section{Summary}\label{sec:6}
The detection capacity of underground DM detection experiments for accelerated light DM largely depends on the scattering processes during particle passage through the Earth (i.e., the Earth-stopping effect). The ES was calculated as the main contributor to the Earth-stopping effect in previous work, but the DM--nucleus scattering could be dominated by IES in the high-kinetic-energy region. Meanwhile, constantly updated experimental data have made precise calculations of DM--nucleus interactions increasingly important. Thus, in this work, we systematically considered and calculated different DM--nucleus scatterings processed via a vector mediator in Earth stopping, including elastic, quasi-elastic, and deep inelastic DM-nucleus scattering. Then, using ``scatter-single'' approximation and two accelerated DM mechanisms (CRDM and BDM), we studied how much these DM--nucleus scatterings affect the flux of DM and the exclusion limit of DM--nucleon spin-independent cross-section in the Xenon1T experiment. Finally, the conclusions of this work are summarized as follows:
\begin{itemize}
    \item The dominant process of DM--nucleus scattering, when mediated by a vector mediator, exhibits a similar trend to that observed in the case of a scalar mediator discussed previously~\cite{Su:2022wpj}. Specifically, for a heavy mediator with a mass of $m_{\gamma^{\prime}} \gtrsim 1$ GeV, the IES process becomes dominant in the high-kinetic-energy region ($T_{\chi} \gtrsim \mathcal{O}(100)$ MeV). This indicates that the nature of the mediator, whether it is a scalar or vector particle, influences the dominant scattering process of DM--nucleus interactions at different energy regimes.

    \item For CRDM particles with masses below approximately 50 MeV, the contribution of inelastic DM--nucleus scattering to their interactions is negligible. This is primarily due to the transfer momentum effect in elastic scattering, which already attenuates a significant portion of the DM particles. As the mass of CRDM particles increases, the transfer momentum effect weakens, and the significance of IES becomes more pronounced. The inclusion of IES in the analysis led to a noticeable difference in the upper bounds of the exclusion limit, which is about a factor of two before and after considering IES.
    \item For BDM $\chi_B$ with $m_{\chi_A} \gtrsim 1 $ GeV, the unique energy spectrum of its flux makes IES very important in direct detection. The difference in the upper bound before and after considering IES became more pronounced as the $m_{\chi_A}$ increased, reaching a gap of several tens of times when $m_{\chi_A}=8$ GeV.
\end{itemize}

The above conclusions are given under ``single-scatter'' approximation. Although this approximation underestimates the number of DM particles reaching the detector and neglects the shift in the energy spectrum due to energy loss, it is enough to show the significance of IES in the direct detection of accelerated DM. A more accurate calculation can be achieved by Monte Carlo simulation of the trajectory of DM on Earth.

\section{Acknowledgments}
We are grateful to Artur M. Ankowski for a useful discussion. This work is supported by the National Natural Science Foundation of China (NNSFC) under grant No. 12275134, 12275232, and 12335005.

\bibliography{refs}

\begin{thebibliography}{102}
\expandafter\ifx\csname natexlab\endcsname\relax\def\natexlab#1{#1}\fi
\expandafter\ifx\csname bibnamefont\endcsname\relax
  \def\bibnamefont#1{#1}\fi
\expandafter\ifx\csname bibfnamefont\endcsname\relax
  \def\bibfnamefont#1{#1}\fi
\expandafter\ifx\csname citenamefont\endcsname\relax
  \def\citenamefont#1{#1}\fi
\expandafter\ifx\csname url\endcsname\relax
  \def\url#1{\texttt{#1}}\fi
\expandafter\ifx\csname urlprefix\endcsname\relax\def\urlprefix{URL }\fi
\providecommand{\bibinfo}[2]{#2}
\providecommand{\eprint}[2][]{\url{#2}}

\bibitem[{\citenamefont{Workman et~al.}(2022)}]{ParticleDataGroup:2022pth}
\bibinfo{author}{\bibfnamefont{R.~L.} \bibnamefont{Workman}}
  \bibnamefont{et~al.} (\bibinfo{collaboration}{Particle Data Group}),
  \bibinfo{journal}{PTEP} \textbf{\bibinfo{volume}{2022}},
  \bibinfo{pages}{083C01} (\bibinfo{year}{2022}).

\bibitem[{\citenamefont{Lee and Weinberg}(1977)}]{Lee:1977ua}
\bibinfo{author}{\bibfnamefont{B.~W.} \bibnamefont{Lee}} \bibnamefont{and}
  \bibinfo{author}{\bibfnamefont{S.}~\bibnamefont{Weinberg}},
  \bibinfo{journal}{Phys. Rev. Lett.} \textbf{\bibinfo{volume}{39}},
  \bibinfo{pages}{165} (\bibinfo{year}{1977}).

\bibitem[{\citenamefont{Jungman et~al.}(1996)\citenamefont{Jungman,
  Kamionkowski, and Griest}}]{Jungman:1995df}
\bibinfo{author}{\bibfnamefont{G.}~\bibnamefont{Jungman}},
  \bibinfo{author}{\bibfnamefont{M.}~\bibnamefont{Kamionkowski}},
  \bibnamefont{and} \bibinfo{author}{\bibfnamefont{K.}~\bibnamefont{Griest}},
  \bibinfo{journal}{Phys. Rept.} \textbf{\bibinfo{volume}{267}},
  \bibinfo{pages}{195} (\bibinfo{year}{1996}), \eprint{hep-ph/9506380}.

\bibitem[{\citenamefont{Meng et~al.}(2021)}]{PandaX-4T:2021bab}
\bibinfo{author}{\bibfnamefont{Y.}~\bibnamefont{Meng}} \bibnamefont{et~al.}
  (\bibinfo{collaboration}{PandaX-4T}), \bibinfo{journal}{Phys. Rev. Lett.}
  \textbf{\bibinfo{volume}{127}}, \bibinfo{pages}{261802}
  (\bibinfo{year}{2021}), \eprint{2107.13438}.

\bibitem[{\citenamefont{Aalbers et~al.}(2022)}]{LZ:2022ufs}
\bibinfo{author}{\bibfnamefont{J.}~\bibnamefont{Aalbers}} \bibnamefont{et~al.}
  (\bibinfo{collaboration}{LZ}) (\bibinfo{year}{2022}), \eprint{2207.03764}.

\bibitem[{\citenamefont{Aprile et~al.}(2023)}]{XENON:2023sxq}
\bibinfo{author}{\bibfnamefont{E.}~\bibnamefont{Aprile}} \bibnamefont{et~al.}
  (\bibinfo{collaboration}{XENON}) (\bibinfo{year}{2023}), \eprint{2303.14729}.

\bibitem[{\citenamefont{Essig et~al.}(2012)\citenamefont{Essig, Mardon, and
  Volansky}}]{Essig:2011nj}
\bibinfo{author}{\bibfnamefont{R.}~\bibnamefont{Essig}},
  \bibinfo{author}{\bibfnamefont{J.}~\bibnamefont{Mardon}}, \bibnamefont{and}
  \bibinfo{author}{\bibfnamefont{T.}~\bibnamefont{Volansky}},
  \bibinfo{journal}{Phys. Rev. D} \textbf{\bibinfo{volume}{85}},
  \bibinfo{pages}{076007} (\bibinfo{year}{2012}), \eprint{1108.5383}.

\bibitem[{\citenamefont{Hochberg
  et~al.}(2016{\natexlab{a}})\citenamefont{Hochberg, Zhao, and
  Zurek}}]{Hochberg:2015pha}
\bibinfo{author}{\bibfnamefont{Y.}~\bibnamefont{Hochberg}},
  \bibinfo{author}{\bibfnamefont{Y.}~\bibnamefont{Zhao}}, \bibnamefont{and}
  \bibinfo{author}{\bibfnamefont{K.~M.} \bibnamefont{Zurek}},
  \bibinfo{journal}{Phys. Rev. Lett.} \textbf{\bibinfo{volume}{116}},
  \bibinfo{pages}{011301} (\bibinfo{year}{2016}{\natexlab{a}}),
  \eprint{1504.07237}.

\bibitem[{\citenamefont{Essig et~al.}(2016)\citenamefont{Essig,
  Fernandez-Serra, Mardon, Soto, Volansky, and Yu}}]{Essig:2015cda}
\bibinfo{author}{\bibfnamefont{R.}~\bibnamefont{Essig}},
  \bibinfo{author}{\bibfnamefont{M.}~\bibnamefont{Fernandez-Serra}},
  \bibinfo{author}{\bibfnamefont{J.}~\bibnamefont{Mardon}},
  \bibinfo{author}{\bibfnamefont{A.}~\bibnamefont{Soto}},
  \bibinfo{author}{\bibfnamefont{T.}~\bibnamefont{Volansky}}, \bibnamefont{and}
  \bibinfo{author}{\bibfnamefont{T.-T.} \bibnamefont{Yu}},
  \bibinfo{journal}{JHEP} \textbf{\bibinfo{volume}{05}}, \bibinfo{pages}{046}
  (\bibinfo{year}{2016}), \eprint{1509.01598}.

\bibitem[{\citenamefont{Essig et~al.}(2017)\citenamefont{Essig, Volansky, and
  Yu}}]{Essig:2017kqs}
\bibinfo{author}{\bibfnamefont{R.}~\bibnamefont{Essig}},
  \bibinfo{author}{\bibfnamefont{T.}~\bibnamefont{Volansky}}, \bibnamefont{and}
  \bibinfo{author}{\bibfnamefont{T.-T.} \bibnamefont{Yu}},
  \bibinfo{journal}{Phys. Rev. D} \textbf{\bibinfo{volume}{96}},
  \bibinfo{pages}{043017} (\bibinfo{year}{2017}), \eprint{1703.00910}.

\bibitem[{\citenamefont{Knapen et~al.}(2017)\citenamefont{Knapen, Lin, and
  Zurek}}]{Knapen:2017xzo}
\bibinfo{author}{\bibfnamefont{S.}~\bibnamefont{Knapen}},
  \bibinfo{author}{\bibfnamefont{T.}~\bibnamefont{Lin}}, \bibnamefont{and}
  \bibinfo{author}{\bibfnamefont{K.~M.} \bibnamefont{Zurek}},
  \bibinfo{journal}{Phys. Rev. D} \textbf{\bibinfo{volume}{96}},
  \bibinfo{pages}{115021} (\bibinfo{year}{2017}), \eprint{1709.07882}.

\bibitem[{\citenamefont{Bertone and Tait}(2018)}]{Bertone:2018krk}
\bibinfo{author}{\bibfnamefont{G.}~\bibnamefont{Bertone}} \bibnamefont{and}
  \bibinfo{author}{\bibfnamefont{M.}~\bibnamefont{Tait}, \bibfnamefont{Tim}},
  \bibinfo{journal}{Nature} \textbf{\bibinfo{volume}{562}}, \bibinfo{pages}{51}
  (\bibinfo{year}{2018}), \eprint{1810.01668}.

\bibitem[{\citenamefont{D'Agnolo et~al.}(2018)\citenamefont{D'Agnolo, Mondino,
  Ruderman, and Wang}}]{DAgnolo:2018wcn}
\bibinfo{author}{\bibfnamefont{R.~T.} \bibnamefont{D'Agnolo}},
  \bibinfo{author}{\bibfnamefont{C.}~\bibnamefont{Mondino}},
  \bibinfo{author}{\bibfnamefont{J.~T.} \bibnamefont{Ruderman}},
  \bibnamefont{and} \bibinfo{author}{\bibfnamefont{P.-J.} \bibnamefont{Wang}},
  \bibinfo{journal}{JHEP} \textbf{\bibinfo{volume}{08}}, \bibinfo{pages}{079}
  (\bibinfo{year}{2018}), \eprint{1803.02901}.

\bibitem[{\citenamefont{Yang et~al.}(2021)}]{PandaX-II:2021lap}
\bibinfo{author}{\bibfnamefont{J.}~\bibnamefont{Yang}} \bibnamefont{et~al.}
  (\bibinfo{collaboration}{PandaX-II}), \bibinfo{journal}{Sci. China Phys.
  Mech. Astron.} \textbf{\bibinfo{volume}{64}}, \bibinfo{pages}{111062}
  (\bibinfo{year}{2021}), \eprint{2104.14724}.

\bibitem[{\citenamefont{Su et~al.}(2022)\citenamefont{Su, Wu, and
  Zhu}}]{Su:2021jvk}
\bibinfo{author}{\bibfnamefont{L.}~\bibnamefont{Su}},
  \bibinfo{author}{\bibfnamefont{L.}~\bibnamefont{Wu}}, \bibnamefont{and}
  \bibinfo{author}{\bibfnamefont{B.}~\bibnamefont{Zhu}},
  \bibinfo{journal}{Phys. Rev. D} \textbf{\bibinfo{volume}{105}},
  \bibinfo{pages}{055021} (\bibinfo{year}{2022}), \eprint{2105.06326}.

\bibitem[{\citenamefont{Calabrese
  et~al.}(2022{\natexlab{a}})\citenamefont{Calabrese, Chianese, Fiorillo, and
  Saviano}}]{Calabrese:2021src}
\bibinfo{author}{\bibfnamefont{R.}~\bibnamefont{Calabrese}},
  \bibinfo{author}{\bibfnamefont{M.}~\bibnamefont{Chianese}},
  \bibinfo{author}{\bibfnamefont{D.~F.~G.} \bibnamefont{Fiorillo}},
  \bibnamefont{and} \bibinfo{author}{\bibfnamefont{N.}~\bibnamefont{Saviano}},
  \bibinfo{journal}{Phys. Rev. D} \textbf{\bibinfo{volume}{105}},
  \bibinfo{pages}{L021302} (\bibinfo{year}{2022}{\natexlab{a}}),
  \eprint{2107.13001}.

\bibitem[{\citenamefont{Elor et~al.}(2023)\citenamefont{Elor, McGehee, and
  Pierce}}]{Elor:2021swj}
\bibinfo{author}{\bibfnamefont{G.}~\bibnamefont{Elor}},
  \bibinfo{author}{\bibfnamefont{R.}~\bibnamefont{McGehee}}, \bibnamefont{and}
  \bibinfo{author}{\bibfnamefont{A.}~\bibnamefont{Pierce}},
  \bibinfo{journal}{Phys. Rev. Lett.} \textbf{\bibinfo{volume}{130}},
  \bibinfo{pages}{031803} (\bibinfo{year}{2023}), \eprint{2112.03920}.

\bibitem[{\citenamefont{Yang et~al.}(2022)}]{Yang:2022eaq}
\bibinfo{author}{\bibfnamefont{H.-B.} \bibnamefont{Yang}} \bibnamefont{et~al.},
  \bibinfo{journal}{Nucl. Sci. Tech.} \textbf{\bibinfo{volume}{33}},
  \bibinfo{pages}{65} (\bibinfo{year}{2022}).

\bibitem[{\citenamefont{Calabrese
  et~al.}(2022{\natexlab{b}})\citenamefont{Calabrese, Chianese, Fiorillo, and
  Saviano}}]{Calabrese:2022rfa}
\bibinfo{author}{\bibfnamefont{R.}~\bibnamefont{Calabrese}},
  \bibinfo{author}{\bibfnamefont{M.}~\bibnamefont{Chianese}},
  \bibinfo{author}{\bibfnamefont{D.~F.~G.} \bibnamefont{Fiorillo}},
  \bibnamefont{and} \bibinfo{author}{\bibfnamefont{N.}~\bibnamefont{Saviano}},
  \bibinfo{journal}{Phys. Rev. D} \textbf{\bibinfo{volume}{105}},
  \bibinfo{pages}{103024} (\bibinfo{year}{2022}{\natexlab{b}}),
  \eprint{2203.17093}.

\bibitem[{\citenamefont{Ambrosone et~al.}(2022)\citenamefont{Ambrosone,
  Chianese, Fiorillo, Marinelli, and Miele}}]{Ambrosone:2022mvk}
\bibinfo{author}{\bibfnamefont{A.}~\bibnamefont{Ambrosone}},
  \bibinfo{author}{\bibfnamefont{M.}~\bibnamefont{Chianese}},
  \bibinfo{author}{\bibfnamefont{D.~F.~G.} \bibnamefont{Fiorillo}},
  \bibinfo{author}{\bibfnamefont{A.}~\bibnamefont{Marinelli}},
  \bibnamefont{and} \bibinfo{author}{\bibfnamefont{G.}~\bibnamefont{Miele}}
  (\bibinfo{year}{2022}), \eprint{2210.05685}.

\bibitem[{\citenamefont{Graham et~al.}(2012)\citenamefont{Graham, Kaplan,
  Rajendran, and Walters}}]{Graham:2012su}
\bibinfo{author}{\bibfnamefont{P.~W.} \bibnamefont{Graham}},
  \bibinfo{author}{\bibfnamefont{D.~E.} \bibnamefont{Kaplan}},
  \bibinfo{author}{\bibfnamefont{S.}~\bibnamefont{Rajendran}},
  \bibnamefont{and} \bibinfo{author}{\bibfnamefont{M.~T.}
  \bibnamefont{Walters}}, \bibinfo{journal}{Phys. Dark Univ.}
  \textbf{\bibinfo{volume}{1}}, \bibinfo{pages}{32} (\bibinfo{year}{2012}),
  \eprint{1203.2531}.

\bibitem[{\citenamefont{Hochberg et~al.}(2017)\citenamefont{Hochberg, Lin, and
  Zurek}}]{Hochberg:2016sqx}
\bibinfo{author}{\bibfnamefont{Y.}~\bibnamefont{Hochberg}},
  \bibinfo{author}{\bibfnamefont{T.}~\bibnamefont{Lin}}, \bibnamefont{and}
  \bibinfo{author}{\bibfnamefont{K.~M.} \bibnamefont{Zurek}},
  \bibinfo{journal}{Phys. Rev. D} \textbf{\bibinfo{volume}{95}},
  \bibinfo{pages}{023013} (\bibinfo{year}{2017}), \eprint{1608.01994}.

\bibitem[{\citenamefont{Bloch et~al.}(2017)\citenamefont{Bloch, Essig, Tobioka,
  Volansky, and Yu}}]{Bloch:2016sjj}
\bibinfo{author}{\bibfnamefont{I.~M.} \bibnamefont{Bloch}},
  \bibinfo{author}{\bibfnamefont{R.}~\bibnamefont{Essig}},
  \bibinfo{author}{\bibfnamefont{K.}~\bibnamefont{Tobioka}},
  \bibinfo{author}{\bibfnamefont{T.}~\bibnamefont{Volansky}}, \bibnamefont{and}
  \bibinfo{author}{\bibfnamefont{T.-T.} \bibnamefont{Yu}},
  \bibinfo{journal}{JHEP} \textbf{\bibinfo{volume}{06}}, \bibinfo{pages}{087}
  (\bibinfo{year}{2017}), \eprint{1608.02123}.

\bibitem[{\citenamefont{Li et~al.}(2022)}]{Li:2022fho}
\bibinfo{author}{\bibfnamefont{R.}~\bibnamefont{Li}} \bibnamefont{et~al.},
  \bibinfo{journal}{Nucl. Sci. Tech.} \textbf{\bibinfo{volume}{33}},
  \bibinfo{pages}{57} (\bibinfo{year}{2022}), \eprint{2201.02961}.

\bibitem[{\citenamefont{Gu et~al.}(2022)\citenamefont{Gu, Wu, and
  Zhu}}]{Gu:2022vgb}
\bibinfo{author}{\bibfnamefont{Y.}~\bibnamefont{Gu}},
  \bibinfo{author}{\bibfnamefont{L.}~\bibnamefont{Wu}}, \bibnamefont{and}
  \bibinfo{author}{\bibfnamefont{B.}~\bibnamefont{Zhu}},
  \bibinfo{journal}{Phys. Rev. D} \textbf{\bibinfo{volume}{106}},
  \bibinfo{pages}{075004} (\bibinfo{year}{2022}), \eprint{2203.06664}.

\bibitem[{\citenamefont{Jia et~al.}(2022)}]{Jia:2022fri}
\bibinfo{author}{\bibfnamefont{H.-T.} \bibnamefont{Jia}} \bibnamefont{et~al.},
  \bibinfo{journal}{Nucl. Sci. Tech.} \textbf{\bibinfo{volume}{33}},
  \bibinfo{pages}{157} (\bibinfo{year}{2022}).

\bibitem[{\citenamefont{Hochberg
  et~al.}(2016{\natexlab{b}})\citenamefont{Hochberg, Lin, and
  Zurek}}]{Hochberg:2016ajh}
\bibinfo{author}{\bibfnamefont{Y.}~\bibnamefont{Hochberg}},
  \bibinfo{author}{\bibfnamefont{T.}~\bibnamefont{Lin}}, \bibnamefont{and}
  \bibinfo{author}{\bibfnamefont{K.~M.} \bibnamefont{Zurek}},
  \bibinfo{journal}{Phys. Rev. D} \textbf{\bibinfo{volume}{94}},
  \bibinfo{pages}{015019} (\bibinfo{year}{2016}{\natexlab{b}}),
  \eprint{1604.06800}.

\bibitem[{\citenamefont{Hochberg et~al.}(2019)\citenamefont{Hochberg, Charaev,
  Nam, Verma, Colangelo, and Berggren}}]{Hochberg:2019cyy}
\bibinfo{author}{\bibfnamefont{Y.}~\bibnamefont{Hochberg}},
  \bibinfo{author}{\bibfnamefont{I.}~\bibnamefont{Charaev}},
  \bibinfo{author}{\bibfnamefont{S.-W.} \bibnamefont{Nam}},
  \bibinfo{author}{\bibfnamefont{V.}~\bibnamefont{Verma}},
  \bibinfo{author}{\bibfnamefont{M.}~\bibnamefont{Colangelo}},
  \bibnamefont{and} \bibinfo{author}{\bibfnamefont{K.~K.}
  \bibnamefont{Berggren}}, \bibinfo{journal}{Phys. Rev. Lett.}
  \textbf{\bibinfo{volume}{123}}, \bibinfo{pages}{151802}
  (\bibinfo{year}{2019}), \eprint{1903.05101}.

\bibitem[{\citenamefont{Griffin et~al.}(2021)\citenamefont{Griffin, Hochberg,
  Inzani, Kurinsky, Lin, and Chin}}]{Griffin:2020lgd}
\bibinfo{author}{\bibfnamefont{S.~M.} \bibnamefont{Griffin}},
  \bibinfo{author}{\bibfnamefont{Y.}~\bibnamefont{Hochberg}},
  \bibinfo{author}{\bibfnamefont{K.}~\bibnamefont{Inzani}},
  \bibinfo{author}{\bibfnamefont{N.}~\bibnamefont{Kurinsky}},
  \bibinfo{author}{\bibfnamefont{T.}~\bibnamefont{Lin}}, \bibnamefont{and}
  \bibinfo{author}{\bibfnamefont{T.}~\bibnamefont{Chin}},
  \bibinfo{journal}{Phys. Rev. D} \textbf{\bibinfo{volume}{103}},
  \bibinfo{pages}{075002} (\bibinfo{year}{2021}), \eprint{2008.08560}.

\bibitem[{\citenamefont{Hochberg et~al.}(2021)\citenamefont{Hochberg, Lehmann,
  Charaev, Chiles, Colangelo, Nam, and Berggren}}]{Hochberg:2021yud}
\bibinfo{author}{\bibfnamefont{Y.}~\bibnamefont{Hochberg}},
  \bibinfo{author}{\bibfnamefont{B.~V.} \bibnamefont{Lehmann}},
  \bibinfo{author}{\bibfnamefont{I.}~\bibnamefont{Charaev}},
  \bibinfo{author}{\bibfnamefont{J.}~\bibnamefont{Chiles}},
  \bibinfo{author}{\bibfnamefont{M.}~\bibnamefont{Colangelo}},
  \bibinfo{author}{\bibfnamefont{S.~W.} \bibnamefont{Nam}}, \bibnamefont{and}
  \bibinfo{author}{\bibfnamefont{K.~K.} \bibnamefont{Berggren}}
  (\bibinfo{year}{2021}), \eprint{2110.01586}.

\bibitem[{\citenamefont{Kouvaris and Pradler}(2017)}]{Kouvaris:2016afs}
\bibinfo{author}{\bibfnamefont{C.}~\bibnamefont{Kouvaris}} \bibnamefont{and}
  \bibinfo{author}{\bibfnamefont{J.}~\bibnamefont{Pradler}},
  \bibinfo{journal}{Phys. Rev. Lett.} \textbf{\bibinfo{volume}{118}},
  \bibinfo{pages}{031803} (\bibinfo{year}{2017}), \eprint{1607.01789}.

\bibitem[{\citenamefont{McCabe}(2017)}]{McCabe:2017rln}
\bibinfo{author}{\bibfnamefont{C.}~\bibnamefont{McCabe}},
  \bibinfo{journal}{Phys. Rev. D} \textbf{\bibinfo{volume}{96}},
  \bibinfo{pages}{043010} (\bibinfo{year}{2017}), \eprint{1702.04730}.

\bibitem[{\citenamefont{Aprile et~al.}(2019)}]{XENON:2019zpr}
\bibinfo{author}{\bibfnamefont{E.}~\bibnamefont{Aprile}} \bibnamefont{et~al.}
  (\bibinfo{collaboration}{XENON}), \bibinfo{journal}{Phys. Rev. Lett.}
  \textbf{\bibinfo{volume}{123}}, \bibinfo{pages}{241803}
  (\bibinfo{year}{2019}), \eprint{1907.12771}.

\bibitem[{\citenamefont{Grilli~di Cortona et~al.}(2020)\citenamefont{Grilli~di
  Cortona, Messina, and Piacentini}}]{GrillidiCortona:2020owp}
\bibinfo{author}{\bibfnamefont{G.}~\bibnamefont{Grilli~di Cortona}},
  \bibinfo{author}{\bibfnamefont{A.}~\bibnamefont{Messina}}, \bibnamefont{and}
  \bibinfo{author}{\bibfnamefont{S.}~\bibnamefont{Piacentini}},
  \bibinfo{journal}{JHEP} \textbf{\bibinfo{volume}{11}}, \bibinfo{pages}{034}
  (\bibinfo{year}{2020}), \eprint{2006.02453}.

\bibitem[{\citenamefont{Vergados and Ejiri}(2005)}]{Vergados:2005dpd}
\bibinfo{author}{\bibfnamefont{J.~D.} \bibnamefont{Vergados}} \bibnamefont{and}
  \bibinfo{author}{\bibfnamefont{H.}~\bibnamefont{Ejiri}},
  \bibinfo{journal}{Phys. Lett. B} \textbf{\bibinfo{volume}{606}},
  \bibinfo{pages}{313} (\bibinfo{year}{2005}), \eprint{hep-ph/0401151}.

\bibitem[{\citenamefont{Moustakidis et~al.}(2005)\citenamefont{Moustakidis,
  Vergados, and Ejiri}}]{Moustakidis:2005gx}
\bibinfo{author}{\bibfnamefont{C.~C.} \bibnamefont{Moustakidis}},
  \bibinfo{author}{\bibfnamefont{J.~D.} \bibnamefont{Vergados}},
  \bibnamefont{and} \bibinfo{author}{\bibfnamefont{H.}~\bibnamefont{Ejiri}},
  \bibinfo{journal}{Nucl. Phys. B} \textbf{\bibinfo{volume}{727}},
  \bibinfo{pages}{406} (\bibinfo{year}{2005}), \eprint{hep-ph/0507123}.

\bibitem[{\citenamefont{Ibe et~al.}(2018)\citenamefont{Ibe, Nakano, Shoji, and
  Suzuki}}]{Ibe:2017yqa}
\bibinfo{author}{\bibfnamefont{M.}~\bibnamefont{Ibe}},
  \bibinfo{author}{\bibfnamefont{W.}~\bibnamefont{Nakano}},
  \bibinfo{author}{\bibfnamefont{Y.}~\bibnamefont{Shoji}}, \bibnamefont{and}
  \bibinfo{author}{\bibfnamefont{K.}~\bibnamefont{Suzuki}},
  \bibinfo{journal}{JHEP} \textbf{\bibinfo{volume}{03}}, \bibinfo{pages}{194}
  (\bibinfo{year}{2018}), \eprint{1707.07258}.

\bibitem[{\citenamefont{Dolan et~al.}(2018)\citenamefont{Dolan, Kahlhoefer, and
  McCabe}}]{Dolan:2017xbu}
\bibinfo{author}{\bibfnamefont{M.~J.} \bibnamefont{Dolan}},
  \bibinfo{author}{\bibfnamefont{F.}~\bibnamefont{Kahlhoefer}},
  \bibnamefont{and} \bibinfo{author}{\bibfnamefont{C.}~\bibnamefont{McCabe}},
  \bibinfo{journal}{Phys. Rev. Lett.} \textbf{\bibinfo{volume}{121}},
  \bibinfo{pages}{101801} (\bibinfo{year}{2018}), \eprint{1711.09906}.

\bibitem[{\citenamefont{Baxter et~al.}(2020)\citenamefont{Baxter, Kahn, and
  Krnjaic}}]{Baxter:2019pnz}
\bibinfo{author}{\bibfnamefont{D.}~\bibnamefont{Baxter}},
  \bibinfo{author}{\bibfnamefont{Y.}~\bibnamefont{Kahn}}, \bibnamefont{and}
  \bibinfo{author}{\bibfnamefont{G.}~\bibnamefont{Krnjaic}},
  \bibinfo{journal}{Phys. Rev. D} \textbf{\bibinfo{volume}{101}},
  \bibinfo{pages}{076014} (\bibinfo{year}{2020}), \eprint{1908.00012}.

\bibitem[{\citenamefont{Essig et~al.}(2020)\citenamefont{Essig, Pradler,
  Sholapurkar, and Yu}}]{Essig:2019xkx}
\bibinfo{author}{\bibfnamefont{R.}~\bibnamefont{Essig}},
  \bibinfo{author}{\bibfnamefont{J.}~\bibnamefont{Pradler}},
  \bibinfo{author}{\bibfnamefont{M.}~\bibnamefont{Sholapurkar}},
  \bibnamefont{and} \bibinfo{author}{\bibfnamefont{T.-T.} \bibnamefont{Yu}},
  \bibinfo{journal}{Phys. Rev. Lett.} \textbf{\bibinfo{volume}{124}},
  \bibinfo{pages}{021801} (\bibinfo{year}{2020}), \eprint{1908.10881}.

\bibitem[{\citenamefont{Bell et~al.}(2020)\citenamefont{Bell, Dent, Newstead,
  Sabharwal, and Weiler}}]{Bell:2019egg}
\bibinfo{author}{\bibfnamefont{N.~F.} \bibnamefont{Bell}},
  \bibinfo{author}{\bibfnamefont{J.~B.} \bibnamefont{Dent}},
  \bibinfo{author}{\bibfnamefont{J.~L.} \bibnamefont{Newstead}},
  \bibinfo{author}{\bibfnamefont{S.}~\bibnamefont{Sabharwal}},
  \bibnamefont{and} \bibinfo{author}{\bibfnamefont{T.~J.}
  \bibnamefont{Weiler}}, \bibinfo{journal}{Phys. Rev. D}
  \textbf{\bibinfo{volume}{101}}, \bibinfo{pages}{015012}
  (\bibinfo{year}{2020}), \eprint{1905.00046}.

\bibitem[{\citenamefont{Liang et~al.}(2020)\citenamefont{Liang, Zhang, Zheng,
  and Zhang}}]{Liang:2019nnx}
\bibinfo{author}{\bibfnamefont{Z.-L.} \bibnamefont{Liang}},
  \bibinfo{author}{\bibfnamefont{L.}~\bibnamefont{Zhang}},
  \bibinfo{author}{\bibfnamefont{F.}~\bibnamefont{Zheng}}, \bibnamefont{and}
  \bibinfo{author}{\bibfnamefont{P.}~\bibnamefont{Zhang}},
  \bibinfo{journal}{Phys. Rev. D} \textbf{\bibinfo{volume}{102}},
  \bibinfo{pages}{043007} (\bibinfo{year}{2020}), \eprint{1912.13484}.

\bibitem[{\citenamefont{Nakamura et~al.}(2020)\citenamefont{Nakamura, Miuchi,
  Kazama, Shoji, Ibe, and Nakano}}]{Nakamura:2020kex}
\bibinfo{author}{\bibfnamefont{K.~D.} \bibnamefont{Nakamura}},
  \bibinfo{author}{\bibfnamefont{K.}~\bibnamefont{Miuchi}},
  \bibinfo{author}{\bibfnamefont{S.}~\bibnamefont{Kazama}},
  \bibinfo{author}{\bibfnamefont{Y.}~\bibnamefont{Shoji}},
  \bibinfo{author}{\bibfnamefont{M.}~\bibnamefont{Ibe}}, \bibnamefont{and}
  \bibinfo{author}{\bibfnamefont{W.}~\bibnamefont{Nakano}}
  (\bibinfo{year}{2020}), \eprint{2009.05939}.

\bibitem[{\citenamefont{Liang et~al.}(2021)\citenamefont{Liang, Mo, Zheng, and
  Zhang}}]{Liang:2020ryg}
\bibinfo{author}{\bibfnamefont{Z.-L.} \bibnamefont{Liang}},
  \bibinfo{author}{\bibfnamefont{C.}~\bibnamefont{Mo}},
  \bibinfo{author}{\bibfnamefont{F.}~\bibnamefont{Zheng}}, \bibnamefont{and}
  \bibinfo{author}{\bibfnamefont{P.}~\bibnamefont{Zhang}},
  \bibinfo{journal}{Phys. Rev. D} \textbf{\bibinfo{volume}{104}},
  \bibinfo{pages}{056009} (\bibinfo{year}{2021}), \eprint{2011.13352}.

\bibitem[{\citenamefont{Flambaum et~al.}(2023)\citenamefont{Flambaum, Su, Wu,
  and Zhu}}]{Flambaum:2020xxo}
\bibinfo{author}{\bibfnamefont{V.~V.} \bibnamefont{Flambaum}},
  \bibinfo{author}{\bibfnamefont{L.}~\bibnamefont{Su}},
  \bibinfo{author}{\bibfnamefont{L.}~\bibnamefont{Wu}}, \bibnamefont{and}
  \bibinfo{author}{\bibfnamefont{B.}~\bibnamefont{Zhu}}, \bibinfo{journal}{Sci.
  China Phys. Mech. Astron.} \textbf{\bibinfo{volume}{66}},
  \bibinfo{pages}{271011} (\bibinfo{year}{2023}), \eprint{2012.09751}.

\bibitem[{\citenamefont{Acevedo et~al.}(2022)\citenamefont{Acevedo, Bramante,
  and Goodman}}]{Acevedo:2021kly}
\bibinfo{author}{\bibfnamefont{J.~F.} \bibnamefont{Acevedo}},
  \bibinfo{author}{\bibfnamefont{J.}~\bibnamefont{Bramante}}, \bibnamefont{and}
  \bibinfo{author}{\bibfnamefont{A.}~\bibnamefont{Goodman}},
  \bibinfo{journal}{Phys. Rev. D} \textbf{\bibinfo{volume}{105}},
  \bibinfo{pages}{023012} (\bibinfo{year}{2022}), \eprint{2108.10889}.

\bibitem[{\citenamefont{Wang et~al.}(2022)\citenamefont{Wang, Wu, Wu, and
  Zhu}}]{Wang:2021oha}
\bibinfo{author}{\bibfnamefont{W.}~\bibnamefont{Wang}},
  \bibinfo{author}{\bibfnamefont{K.-Y.} \bibnamefont{Wu}},
  \bibinfo{author}{\bibfnamefont{L.}~\bibnamefont{Wu}}, \bibnamefont{and}
  \bibinfo{author}{\bibfnamefont{B.}~\bibnamefont{Zhu}},
  \bibinfo{journal}{Nucl. Phys. B} \textbf{\bibinfo{volume}{983}},
  \bibinfo{pages}{115907} (\bibinfo{year}{2022}), \eprint{2112.06492}.

\bibitem[{\citenamefont{Bell et~al.}(2022)\citenamefont{Bell, Dent, Lang,
  Newstead, and Ritter}}]{Bell:2021ihi}
\bibinfo{author}{\bibfnamefont{N.~F.} \bibnamefont{Bell}},
  \bibinfo{author}{\bibfnamefont{J.~B.} \bibnamefont{Dent}},
  \bibinfo{author}{\bibfnamefont{R.~F.} \bibnamefont{Lang}},
  \bibinfo{author}{\bibfnamefont{J.~L.} \bibnamefont{Newstead}},
  \bibnamefont{and} \bibinfo{author}{\bibfnamefont{A.~C.}
  \bibnamefont{Ritter}}, \bibinfo{journal}{Phys. Rev. D}
  \textbf{\bibinfo{volume}{105}}, \bibinfo{pages}{096015}
  (\bibinfo{year}{2022}), \eprint{2112.08514}.

\bibitem[{\citenamefont{Cox et~al.}(2023)\citenamefont{Cox, Dolan, McCabe, and
  Quiney}}]{Cox:2022ekg}
\bibinfo{author}{\bibfnamefont{P.}~\bibnamefont{Cox}},
  \bibinfo{author}{\bibfnamefont{M.~J.} \bibnamefont{Dolan}},
  \bibinfo{author}{\bibfnamefont{C.}~\bibnamefont{McCabe}}, \bibnamefont{and}
  \bibinfo{author}{\bibfnamefont{H.~M.} \bibnamefont{Quiney}},
  \bibinfo{journal}{Phys. Rev. D} \textbf{\bibinfo{volume}{107}},
  \bibinfo{pages}{035032} (\bibinfo{year}{2023}), \eprint{2208.12222}.

\bibitem[{\citenamefont{Berghaus et~al.}(2022)\citenamefont{Berghaus, Esposito,
  Essig, and Sholapurkar}}]{Berghaus:2022pbu}
\bibinfo{author}{\bibfnamefont{K.~V.} \bibnamefont{Berghaus}},
  \bibinfo{author}{\bibfnamefont{A.}~\bibnamefont{Esposito}},
  \bibinfo{author}{\bibfnamefont{R.}~\bibnamefont{Essig}}, \bibnamefont{and}
  \bibinfo{author}{\bibfnamefont{M.}~\bibnamefont{Sholapurkar}}
  (\bibinfo{year}{2022}), \eprint{2210.06490}.

\bibitem[{\citenamefont{Li et~al.}(2023)\citenamefont{Li, Su, Wu, and
  Zhu}}]{Li:2022acp}
\bibinfo{author}{\bibfnamefont{J.}~\bibnamefont{Li}},
  \bibinfo{author}{\bibfnamefont{L.}~\bibnamefont{Su}},
  \bibinfo{author}{\bibfnamefont{L.}~\bibnamefont{Wu}}, \bibnamefont{and}
  \bibinfo{author}{\bibfnamefont{B.}~\bibnamefont{Zhu}},
  \bibinfo{journal}{JCAP} \textbf{\bibinfo{volume}{04}}, \bibinfo{pages}{020}
  (\bibinfo{year}{2023}), \eprint{2210.15474}.

\bibitem[{\citenamefont{Adams et~al.}(2023)\citenamefont{Adams, Baxter, Day,
  Essig, and Kahn}}]{Adams:2022zvg}
\bibinfo{author}{\bibfnamefont{D.}~\bibnamefont{Adams}},
  \bibinfo{author}{\bibfnamefont{D.}~\bibnamefont{Baxter}},
  \bibinfo{author}{\bibfnamefont{H.}~\bibnamefont{Day}},
  \bibinfo{author}{\bibfnamefont{R.}~\bibnamefont{Essig}}, \bibnamefont{and}
  \bibinfo{author}{\bibfnamefont{Y.}~\bibnamefont{Kahn}},
  \bibinfo{journal}{Phys. Rev. D} \textbf{\bibinfo{volume}{107}},
  \bibinfo{pages}{L041303} (\bibinfo{year}{2023}), \eprint{2210.04917}.

\bibitem[{\citenamefont{Bell et~al.}(2023)\citenamefont{Bell, Cox, Dolan,
  Newstead, and Ritter}}]{Bell:2023uvf}
\bibinfo{author}{\bibfnamefont{N.~F.} \bibnamefont{Bell}},
  \bibinfo{author}{\bibfnamefont{P.}~\bibnamefont{Cox}},
  \bibinfo{author}{\bibfnamefont{M.~J.} \bibnamefont{Dolan}},
  \bibinfo{author}{\bibfnamefont{J.~L.} \bibnamefont{Newstead}},
  \bibnamefont{and} \bibinfo{author}{\bibfnamefont{A.~C.} \bibnamefont{Ritter}}
  (\bibinfo{year}{2023}), \eprint{2305.04690}.

\bibitem[{\citenamefont{Qiao et~al.}(2023)\citenamefont{Qiao, Xia, and
  Zhou}}]{Qiao:2023pbw}
\bibinfo{author}{\bibfnamefont{M.}~\bibnamefont{Qiao}},
  \bibinfo{author}{\bibfnamefont{C.}~\bibnamefont{Xia}}, \bibnamefont{and}
  \bibinfo{author}{\bibfnamefont{Y.-F.} \bibnamefont{Zhou}}
  (\bibinfo{year}{2023}), \eprint{2307.12820}.

\bibitem[{\citenamefont{Bringmann and Pospelov}(2019)}]{Bringmann:2018cvk}
\bibinfo{author}{\bibfnamefont{T.}~\bibnamefont{Bringmann}} \bibnamefont{and}
  \bibinfo{author}{\bibfnamefont{M.}~\bibnamefont{Pospelov}},
  \bibinfo{journal}{Phys. Rev. Lett.} \textbf{\bibinfo{volume}{122}},
  \bibinfo{pages}{171801} (\bibinfo{year}{2019}), \eprint{1810.10543}.

\bibitem[{\citenamefont{Ema et~al.}(2019)\citenamefont{Ema, Sala, and
  Sato}}]{Ema:2018bih}
\bibinfo{author}{\bibfnamefont{Y.}~\bibnamefont{Ema}},
  \bibinfo{author}{\bibfnamefont{F.}~\bibnamefont{Sala}}, \bibnamefont{and}
  \bibinfo{author}{\bibfnamefont{R.}~\bibnamefont{Sato}},
  \bibinfo{journal}{Phys. Rev. Lett.} \textbf{\bibinfo{volume}{122}},
  \bibinfo{pages}{181802} (\bibinfo{year}{2019}), \eprint{1811.00520}.

\bibitem[{\citenamefont{Cappiello and Beacom}(2019)}]{Cappiello:2019qsw}
\bibinfo{author}{\bibfnamefont{C.~V.} \bibnamefont{Cappiello}}
  \bibnamefont{and} \bibinfo{author}{\bibfnamefont{J.~F.}
  \bibnamefont{Beacom}}, \bibinfo{journal}{Phys. Rev. D}
  \textbf{\bibinfo{volume}{100}}, \bibinfo{pages}{103011}
  (\bibinfo{year}{2019}), \bibinfo{note}{[Erratum: Phys.Rev.D 104, 069901
  (2021)]}, \eprint{1906.11283}.

\bibitem[{\citenamefont{Wang et~al.}(2020)\citenamefont{Wang, Wu, Yang, Zhou,
  and Zhu}}]{Wang:2019jtk}
\bibinfo{author}{\bibfnamefont{W.}~\bibnamefont{Wang}},
  \bibinfo{author}{\bibfnamefont{L.}~\bibnamefont{Wu}},
  \bibinfo{author}{\bibfnamefont{J.~M.} \bibnamefont{Yang}},
  \bibinfo{author}{\bibfnamefont{H.}~\bibnamefont{Zhou}}, \bibnamefont{and}
  \bibinfo{author}{\bibfnamefont{B.}~\bibnamefont{Zhu}},
  \bibinfo{journal}{JHEP} \textbf{\bibinfo{volume}{12}}, \bibinfo{pages}{072}
  (\bibinfo{year}{2020}), \bibinfo{note}{[Erratum: JHEP 02, 052 (2021)]},
  \eprint{1912.09904}.

\bibitem[{\citenamefont{Ge et~al.}(2021)\citenamefont{Ge, Liu, Yuan, and
  Zhou}}]{Ge:2020yuf}
\bibinfo{author}{\bibfnamefont{S.-F.} \bibnamefont{Ge}},
  \bibinfo{author}{\bibfnamefont{J.}~\bibnamefont{Liu}},
  \bibinfo{author}{\bibfnamefont{Q.}~\bibnamefont{Yuan}}, \bibnamefont{and}
  \bibinfo{author}{\bibfnamefont{N.}~\bibnamefont{Zhou}},
  \bibinfo{journal}{Phys. Rev. Lett.} \textbf{\bibinfo{volume}{126}},
  \bibinfo{pages}{091804} (\bibinfo{year}{2021}), \eprint{2005.09480}.

\bibitem[{\citenamefont{Xia et~al.}(2021)\citenamefont{Xia, Xu, and
  Zhou}}]{Xia:2020apm}
\bibinfo{author}{\bibfnamefont{C.}~\bibnamefont{Xia}},
  \bibinfo{author}{\bibfnamefont{Y.-H.} \bibnamefont{Xu}}, \bibnamefont{and}
  \bibinfo{author}{\bibfnamefont{Y.-F.} \bibnamefont{Zhou}},
  \bibinfo{journal}{Nucl. Phys. B} \textbf{\bibinfo{volume}{969}},
  \bibinfo{pages}{115470} (\bibinfo{year}{2021}), \eprint{2009.00353}.

\bibitem[{\citenamefont{Ema et~al.}(2021)\citenamefont{Ema, Sala, and
  Sato}}]{Ema:2020ulo}
\bibinfo{author}{\bibfnamefont{Y.}~\bibnamefont{Ema}},
  \bibinfo{author}{\bibfnamefont{F.}~\bibnamefont{Sala}}, \bibnamefont{and}
  \bibinfo{author}{\bibfnamefont{R.}~\bibnamefont{Sato}},
  \bibinfo{journal}{SciPost Phys.} \textbf{\bibinfo{volume}{10}},
  \bibinfo{pages}{072} (\bibinfo{year}{2021}), \eprint{2011.01939}.

\bibitem[{\citenamefont{Bell et~al.}(2021)\citenamefont{Bell, Dent, Dutta,
  Ghosh, Kumar, Newstead, and Shoemaker}}]{Bell:2021xff}
\bibinfo{author}{\bibfnamefont{N.~F.} \bibnamefont{Bell}},
  \bibinfo{author}{\bibfnamefont{J.~B.} \bibnamefont{Dent}},
  \bibinfo{author}{\bibfnamefont{B.}~\bibnamefont{Dutta}},
  \bibinfo{author}{\bibfnamefont{S.}~\bibnamefont{Ghosh}},
  \bibinfo{author}{\bibfnamefont{J.}~\bibnamefont{Kumar}},
  \bibinfo{author}{\bibfnamefont{J.~L.} \bibnamefont{Newstead}},
  \bibnamefont{and} \bibinfo{author}{\bibfnamefont{I.~M.}
  \bibnamefont{Shoemaker}}, \bibinfo{journal}{Phys. Rev. D}
  \textbf{\bibinfo{volume}{104}}, \bibinfo{pages}{076020}
  (\bibinfo{year}{2021}), \eprint{2108.00583}.

\bibitem[{\citenamefont{Feng et~al.}(2022)\citenamefont{Feng, Kang, Lu, Tsai,
  and Zhang}}]{Feng:2021hyz}
\bibinfo{author}{\bibfnamefont{J.-C.} \bibnamefont{Feng}},
  \bibinfo{author}{\bibfnamefont{X.-W.} \bibnamefont{Kang}},
  \bibinfo{author}{\bibfnamefont{C.-T.} \bibnamefont{Lu}},
  \bibinfo{author}{\bibfnamefont{Y.-L.~S.} \bibnamefont{Tsai}},
  \bibnamefont{and} \bibinfo{author}{\bibfnamefont{F.-S.} \bibnamefont{Zhang}},
  \bibinfo{journal}{JHEP} \textbf{\bibinfo{volume}{04}}, \bibinfo{pages}{080}
  (\bibinfo{year}{2022}), \eprint{2110.08863}.

\bibitem[{\citenamefont{Wang et~al.}(2023)\citenamefont{Wang, Wu, Yang, and
  Zhu}}]{Wang:2021nbf}
\bibinfo{author}{\bibfnamefont{W.}~\bibnamefont{Wang}},
  \bibinfo{author}{\bibfnamefont{L.}~\bibnamefont{Wu}},
  \bibinfo{author}{\bibfnamefont{W.-N.} \bibnamefont{Yang}}, \bibnamefont{and}
  \bibinfo{author}{\bibfnamefont{B.}~\bibnamefont{Zhu}},
  \bibinfo{journal}{Phys. Rev. D} \textbf{\bibinfo{volume}{107}},
  \bibinfo{pages}{073002} (\bibinfo{year}{2023}), \eprint{2111.04000}.

\bibitem[{\citenamefont{Cui et~al.}(2022)}]{PandaX-II:2021kai}
\bibinfo{author}{\bibfnamefont{X.}~\bibnamefont{Cui}} \bibnamefont{et~al.}
  (\bibinfo{collaboration}{PandaX-II}), \bibinfo{journal}{Phys. Rev. Lett.}
  \textbf{\bibinfo{volume}{128}}, \bibinfo{pages}{171801}
  (\bibinfo{year}{2022}), \eprint{2112.08957}.

\bibitem[{\citenamefont{Maity and Laha}(2022)}]{Maity:2022exk}
\bibinfo{author}{\bibfnamefont{T.~N.} \bibnamefont{Maity}} \bibnamefont{and}
  \bibinfo{author}{\bibfnamefont{R.}~\bibnamefont{Laha}}
  (\bibinfo{year}{2022}), \eprint{2210.01815}.

\bibitem[{\citenamefont{Nagao et~al.}(2023)\citenamefont{Nagao, Higashino,
  Naka, and Miuchi}}]{Nagao:2022azp}
\bibinfo{author}{\bibfnamefont{K.~I.} \bibnamefont{Nagao}},
  \bibinfo{author}{\bibfnamefont{S.}~\bibnamefont{Higashino}},
  \bibinfo{author}{\bibfnamefont{T.}~\bibnamefont{Naka}}, \bibnamefont{and}
  \bibinfo{author}{\bibfnamefont{K.}~\bibnamefont{Miuchi}},
  \bibinfo{journal}{JCAP} \textbf{\bibinfo{volume}{07}}, \bibinfo{pages}{061}
  (\bibinfo{year}{2023}), \eprint{2211.13399}.

\bibitem[{\citenamefont{Agashe et~al.}(2014)\citenamefont{Agashe, Cui, Necib,
  and Thaler}}]{Agashe:2014yua}
\bibinfo{author}{\bibfnamefont{K.}~\bibnamefont{Agashe}},
  \bibinfo{author}{\bibfnamefont{Y.}~\bibnamefont{Cui}},
  \bibinfo{author}{\bibfnamefont{L.}~\bibnamefont{Necib}}, \bibnamefont{and}
  \bibinfo{author}{\bibfnamefont{J.}~\bibnamefont{Thaler}},
  \bibinfo{journal}{JCAP} \textbf{\bibinfo{volume}{10}}, \bibinfo{pages}{062}
  (\bibinfo{year}{2014}), \eprint{1405.7370}.

\bibitem[{\citenamefont{Berger et~al.}(2015)\citenamefont{Berger, Cui, and
  Zhao}}]{Berger:2014sqa}
\bibinfo{author}{\bibfnamefont{J.}~\bibnamefont{Berger}},
  \bibinfo{author}{\bibfnamefont{Y.}~\bibnamefont{Cui}}, \bibnamefont{and}
  \bibinfo{author}{\bibfnamefont{Y.}~\bibnamefont{Zhao}},
  \bibinfo{journal}{JCAP} \textbf{\bibinfo{volume}{02}}, \bibinfo{pages}{005}
  (\bibinfo{year}{2015}), \eprint{1410.2246}.

\bibitem[{\citenamefont{Agashe et~al.}(2016)\citenamefont{Agashe, Cui, Necib,
  and Thaler}}]{Agashe:2015xkj}
\bibinfo{author}{\bibfnamefont{K.}~\bibnamefont{Agashe}},
  \bibinfo{author}{\bibfnamefont{Y.}~\bibnamefont{Cui}},
  \bibinfo{author}{\bibfnamefont{L.}~\bibnamefont{Necib}}, \bibnamefont{and}
  \bibinfo{author}{\bibfnamefont{J.}~\bibnamefont{Thaler}},
  \bibinfo{journal}{J. Phys. Conf. Ser.} \textbf{\bibinfo{volume}{718}},
  \bibinfo{pages}{042041} (\bibinfo{year}{2016}), \eprint{1512.03782}.

\bibitem[{\citenamefont{Alvey et~al.}(2019)\citenamefont{Alvey, Campos,
  Fairbairn, and You}}]{Alvey:2019zaa}
\bibinfo{author}{\bibfnamefont{J.}~\bibnamefont{Alvey}},
  \bibinfo{author}{\bibfnamefont{M.}~\bibnamefont{Campos}},
  \bibinfo{author}{\bibfnamefont{M.}~\bibnamefont{Fairbairn}},
  \bibnamefont{and} \bibinfo{author}{\bibfnamefont{T.}~\bibnamefont{You}},
  \bibinfo{journal}{Phys. Rev. Lett.} \textbf{\bibinfo{volume}{123}},
  \bibinfo{pages}{261802} (\bibinfo{year}{2019}), \eprint{1905.05776}.

\bibitem[{\citenamefont{Su et~al.}(2020)\citenamefont{Su, Wang, Wu, Yang, and
  Zhu}}]{Su:2020zny}
\bibinfo{author}{\bibfnamefont{L.}~\bibnamefont{Su}},
  \bibinfo{author}{\bibfnamefont{W.}~\bibnamefont{Wang}},
  \bibinfo{author}{\bibfnamefont{L.}~\bibnamefont{Wu}},
  \bibinfo{author}{\bibfnamefont{J.~M.} \bibnamefont{Yang}}, \bibnamefont{and}
  \bibinfo{author}{\bibfnamefont{B.}~\bibnamefont{Zhu}},
  \bibinfo{journal}{Phys. Rev. D} \textbf{\bibinfo{volume}{102}},
  \bibinfo{pages}{115028} (\bibinfo{year}{2020}), \eprint{2006.11837}.

\bibitem[{\citenamefont{Arg\"uelles et~al.}(2022)\citenamefont{Arg\"uelles,
  Mu\~noz, Shoemaker, and Takhistov}}]{Arguelles:2022fqq}
\bibinfo{author}{\bibfnamefont{C.~A.} \bibnamefont{Arg\"uelles}},
  \bibinfo{author}{\bibfnamefont{V.}~\bibnamefont{Mu\~noz}},
  \bibinfo{author}{\bibfnamefont{I.~M.} \bibnamefont{Shoemaker}},
  \bibnamefont{and}
  \bibinfo{author}{\bibfnamefont{V.}~\bibnamefont{Takhistov}},
  \bibinfo{journal}{Phys. Lett. B} \textbf{\bibinfo{volume}{833}},
  \bibinfo{pages}{137363} (\bibinfo{year}{2022}), \eprint{2203.12630}.

\bibitem[{\citenamefont{Darm\'e}(2022)}]{Darme:2022bew}
\bibinfo{author}{\bibfnamefont{L.}~\bibnamefont{Darm\'e}},
  \bibinfo{journal}{Phys. Rev. D} \textbf{\bibinfo{volume}{106}},
  \bibinfo{pages}{055015} (\bibinfo{year}{2022}), \eprint{2205.09773}.

\bibitem[{\citenamefont{Du et~al.}(2022)\citenamefont{Du, Fang, and
  Liu}}]{Du:2022hms}
\bibinfo{author}{\bibfnamefont{M.}~\bibnamefont{Du}},
  \bibinfo{author}{\bibfnamefont{R.}~\bibnamefont{Fang}}, \bibnamefont{and}
  \bibinfo{author}{\bibfnamefont{Z.}~\bibnamefont{Liu}} (\bibinfo{year}{2022}),
  \eprint{2211.11469}.

\bibitem[{\citenamefont{Sieber et~al.}(2023)\citenamefont{Sieber, Kirpichnikov,
  Voronchikhin, Crivelli, Gninenko, Kirsanov, Krasnikov, Molina-Bueno, and
  Sekatskii}}]{Sieber:2023nkq}
\bibinfo{author}{\bibfnamefont{H.}~\bibnamefont{Sieber}},
  \bibinfo{author}{\bibfnamefont{D.~V.} \bibnamefont{Kirpichnikov}},
  \bibinfo{author}{\bibfnamefont{I.~V.} \bibnamefont{Voronchikhin}},
  \bibinfo{author}{\bibfnamefont{P.}~\bibnamefont{Crivelli}},
  \bibinfo{author}{\bibfnamefont{S.~N.} \bibnamefont{Gninenko}},
  \bibinfo{author}{\bibfnamefont{M.~M.} \bibnamefont{Kirsanov}},
  \bibinfo{author}{\bibfnamefont{N.~V.} \bibnamefont{Krasnikov}},
  \bibinfo{author}{\bibfnamefont{L.}~\bibnamefont{Molina-Bueno}},
  \bibnamefont{and} \bibinfo{author}{\bibfnamefont{S.}~\bibnamefont{Sekatskii}}
  (\bibinfo{year}{2023}), \eprint{2305.09015}.

\bibitem[{\citenamefont{Kouvaris and Shoemaker}(2014)}]{Kouvaris:2014lpa}
\bibinfo{author}{\bibfnamefont{C.}~\bibnamefont{Kouvaris}} \bibnamefont{and}
  \bibinfo{author}{\bibfnamefont{I.~M.} \bibnamefont{Shoemaker}},
  \bibinfo{journal}{Phys. Rev. D} \textbf{\bibinfo{volume}{90}},
  \bibinfo{pages}{095011} (\bibinfo{year}{2014}), \eprint{1405.1729}.

\bibitem[{\citenamefont{Foot and Vagnozzi}(2015)}]{Foot:2014osa}
\bibinfo{author}{\bibfnamefont{R.}~\bibnamefont{Foot}} \bibnamefont{and}
  \bibinfo{author}{\bibfnamefont{S.}~\bibnamefont{Vagnozzi}},
  \bibinfo{journal}{Phys. Lett. B} \textbf{\bibinfo{volume}{748}},
  \bibinfo{pages}{61} (\bibinfo{year}{2015}), \eprint{1412.0762}.

\bibitem[{\citenamefont{Kavanagh et~al.}(2017)\citenamefont{Kavanagh, Catena,
  and Kouvaris}}]{Kavanagh:2016pyr}
\bibinfo{author}{\bibfnamefont{B.~J.} \bibnamefont{Kavanagh}},
  \bibinfo{author}{\bibfnamefont{R.}~\bibnamefont{Catena}}, \bibnamefont{and}
  \bibinfo{author}{\bibfnamefont{C.}~\bibnamefont{Kouvaris}},
  \bibinfo{journal}{JCAP} \textbf{\bibinfo{volume}{01}}, \bibinfo{pages}{012}
  (\bibinfo{year}{2017}), \eprint{1611.05453}.

\bibitem[{\citenamefont{Kavanagh}(2018)}]{Kavanagh:2017cru}
\bibinfo{author}{\bibfnamefont{B.~J.} \bibnamefont{Kavanagh}},
  \bibinfo{journal}{Phys. Rev. D} \textbf{\bibinfo{volume}{97}},
  \bibinfo{pages}{123013} (\bibinfo{year}{2018}), \eprint{1712.04901}.

\bibitem[{\citenamefont{Alvey et~al.}(2023)\citenamefont{Alvey, Bringmann, and
  Kolesova}}]{Alvey:2022pad}
\bibinfo{author}{\bibfnamefont{J.}~\bibnamefont{Alvey}},
  \bibinfo{author}{\bibfnamefont{T.}~\bibnamefont{Bringmann}},
  \bibnamefont{and} \bibinfo{author}{\bibfnamefont{H.}~\bibnamefont{Kolesova}},
  \bibinfo{journal}{JHEP} \textbf{\bibinfo{volume}{01}}, \bibinfo{pages}{123}
  (\bibinfo{year}{2023}), \eprint{2209.03360}.

\bibitem[{\citenamefont{Su et~al.}(2023)\citenamefont{Su, Wu, Zhou, and
  Zhu}}]{Su:2022wpj}
\bibinfo{author}{\bibfnamefont{L.}~\bibnamefont{Su}},
  \bibinfo{author}{\bibfnamefont{L.}~\bibnamefont{Wu}},
  \bibinfo{author}{\bibfnamefont{N.}~\bibnamefont{Zhou}}, \bibnamefont{and}
  \bibinfo{author}{\bibfnamefont{B.}~\bibnamefont{Zhu}},
  \bibinfo{journal}{Phys. Rev. D} \textbf{\bibinfo{volume}{108}},
  \bibinfo{pages}{035004} (\bibinfo{year}{2023}), \eprint{2212.02286}.

\bibitem[{\citenamefont{Ning et~al.}(2023)}]{PandaX:2023tfq}
\bibinfo{author}{\bibfnamefont{X.}~\bibnamefont{Ning}} \bibnamefont{et~al.}
  (\bibinfo{collaboration}{PandaX}), \bibinfo{journal}{Phys. Rev. Lett.}
  \textbf{\bibinfo{volume}{131}}, \bibinfo{pages}{041001}
  (\bibinfo{year}{2023}), \eprint{2301.03010}.

\bibitem[{\citenamefont{Arg\"uelles et~al.}(2021)\citenamefont{Arg\"uelles,
  Diaz, Kheirandish, Olivares-Del-Campo, Safa, and
  Vincent}}]{Arguelles:2019ouk}
\bibinfo{author}{\bibfnamefont{C.~A.} \bibnamefont{Arg\"uelles}},
  \bibinfo{author}{\bibfnamefont{A.}~\bibnamefont{Diaz}},
  \bibinfo{author}{\bibfnamefont{A.}~\bibnamefont{Kheirandish}},
  \bibinfo{author}{\bibfnamefont{A.}~\bibnamefont{Olivares-Del-Campo}},
  \bibinfo{author}{\bibfnamefont{I.}~\bibnamefont{Safa}}, \bibnamefont{and}
  \bibinfo{author}{\bibfnamefont{A.~C.} \bibnamefont{Vincent}},
  \bibinfo{journal}{Rev. Mod. Phys.} \textbf{\bibinfo{volume}{93}},
  \bibinfo{pages}{035007} (\bibinfo{year}{2021}), \eprint{1912.09486}.

\bibitem[{\citenamefont{Boschini et~al.}(2017)}]{Boschini:2017fxq}
\bibinfo{author}{\bibfnamefont{M.~J.} \bibnamefont{Boschini}}
  \bibnamefont{et~al.}, \bibinfo{journal}{Astrophys. J.}
  \textbf{\bibinfo{volume}{840}}, \bibinfo{pages}{115} (\bibinfo{year}{2017}),
  \eprint{1704.06337}.

\bibitem[{\citenamefont{Guo et~al.}(2020)\citenamefont{Guo, Tsai, Wu, and
  Yuan}}]{Guo:2020oum}
\bibinfo{author}{\bibfnamefont{G.}~\bibnamefont{Guo}},
  \bibinfo{author}{\bibfnamefont{Y.-L.~S.} \bibnamefont{Tsai}},
  \bibinfo{author}{\bibfnamefont{M.-R.} \bibnamefont{Wu}}, \bibnamefont{and}
  \bibinfo{author}{\bibfnamefont{Q.}~\bibnamefont{Yuan}},
  \bibinfo{journal}{Phys. Rev. D} \textbf{\bibinfo{volume}{102}},
  \bibinfo{pages}{103004} (\bibinfo{year}{2020}), \eprint{2008.12137}.

\bibitem[{\citenamefont{Fabbrichesi et~al.}(2020)\citenamefont{Fabbrichesi,
  Gabrielli, and Lanfranchi}}]{Fabbrichesi:2020wbt}
\bibinfo{author}{\bibfnamefont{M.}~\bibnamefont{Fabbrichesi}},
  \bibinfo{author}{\bibfnamefont{E.}~\bibnamefont{Gabrielli}},
  \bibnamefont{and}
  \bibinfo{author}{\bibfnamefont{G.}~\bibnamefont{Lanfranchi}}
  (\bibinfo{year}{2020}), \eprint{2005.01515}.

\bibitem[{\citenamefont{Dent et~al.}(2015)\citenamefont{Dent, Krauss, Newstead,
  and Sabharwal}}]{Dent:2015zpa}
\bibinfo{author}{\bibfnamefont{J.~B.} \bibnamefont{Dent}},
  \bibinfo{author}{\bibfnamefont{L.~M.} \bibnamefont{Krauss}},
  \bibinfo{author}{\bibfnamefont{J.~L.} \bibnamefont{Newstead}},
  \bibnamefont{and}
  \bibinfo{author}{\bibfnamefont{S.}~\bibnamefont{Sabharwal}},
  \bibinfo{journal}{Phys. Rev. D} \textbf{\bibinfo{volume}{92}},
  \bibinfo{pages}{063515} (\bibinfo{year}{2015}), \eprint{1505.03117}.

\bibitem[{\citenamefont{Aristizabal~Sierra
  et~al.}(2018)\citenamefont{Aristizabal~Sierra, De~Romeri, and
  Rojas}}]{AristizabalSierra:2018eqm}
\bibinfo{author}{\bibfnamefont{D.}~\bibnamefont{Aristizabal~Sierra}},
  \bibinfo{author}{\bibfnamefont{V.}~\bibnamefont{De~Romeri}},
  \bibnamefont{and} \bibinfo{author}{\bibfnamefont{N.}~\bibnamefont{Rojas}},
  \bibinfo{journal}{Phys. Rev. D} \textbf{\bibinfo{volume}{98}},
  \bibinfo{pages}{075018} (\bibinfo{year}{2018}), \eprint{1806.07424}.

\bibitem[{\citenamefont{Buckley et~al.}(2015)\citenamefont{Buckley, Ferrando,
  Lloyd, Nordstr\"om, Page, R\"ufenacht, Sch\"onherr, and
  Watt}}]{Buckley:2014ana}
\bibinfo{author}{\bibfnamefont{A.}~\bibnamefont{Buckley}},
  \bibinfo{author}{\bibfnamefont{J.}~\bibnamefont{Ferrando}},
  \bibinfo{author}{\bibfnamefont{S.}~\bibnamefont{Lloyd}},
  \bibinfo{author}{\bibfnamefont{K.}~\bibnamefont{Nordstr\"om}},
  \bibinfo{author}{\bibfnamefont{B.}~\bibnamefont{Page}},
  \bibinfo{author}{\bibfnamefont{M.}~\bibnamefont{R\"ufenacht}},
  \bibinfo{author}{\bibfnamefont{M.}~\bibnamefont{Sch\"onherr}},
  \bibnamefont{and} \bibinfo{author}{\bibfnamefont{G.}~\bibnamefont{Watt}},
  \bibinfo{journal}{Eur. Phys. J. C} \textbf{\bibinfo{volume}{75}},
  \bibinfo{pages}{132} (\bibinfo{year}{2015}), \eprint{1412.7420}.

\bibitem[{\citenamefont{Abdul~Khalek et~al.}(2022)\citenamefont{Abdul~Khalek,
  Gauld, Giani, Nocera, Rabemananjara, and Rojo}}]{AbdulKhalek:2022fyi}
\bibinfo{author}{\bibfnamefont{R.}~\bibnamefont{Abdul~Khalek}},
  \bibinfo{author}{\bibfnamefont{R.}~\bibnamefont{Gauld}},
  \bibinfo{author}{\bibfnamefont{T.}~\bibnamefont{Giani}},
  \bibinfo{author}{\bibfnamefont{E.~R.} \bibnamefont{Nocera}},
  \bibinfo{author}{\bibfnamefont{T.~R.} \bibnamefont{Rabemananjara}},
  \bibnamefont{and} \bibinfo{author}{\bibfnamefont{J.}~\bibnamefont{Rojo}},
  \bibinfo{journal}{Eur. Phys. J. C} \textbf{\bibinfo{volume}{82}},
  \bibinfo{pages}{507} (\bibinfo{year}{2022}), \eprint{2201.12363}.

\bibitem[{\citenamefont{Benhar et~al.}(2005)\citenamefont{Benhar, Farina,
  Nakamura, Sakuda, and Seki}}]{Benhar:2005dj}
\bibinfo{author}{\bibfnamefont{O.}~\bibnamefont{Benhar}},
  \bibinfo{author}{\bibfnamefont{N.}~\bibnamefont{Farina}},
  \bibinfo{author}{\bibfnamefont{H.}~\bibnamefont{Nakamura}},
  \bibinfo{author}{\bibfnamefont{M.}~\bibnamefont{Sakuda}}, \bibnamefont{and}
  \bibinfo{author}{\bibfnamefont{R.}~\bibnamefont{Seki}},
  \bibinfo{journal}{Phys. Rev. D} \textbf{\bibinfo{volume}{72}},
  \bibinfo{pages}{053005} (\bibinfo{year}{2005}), \eprint{hep-ph/0506116}.

\bibitem[{\citenamefont{Ankowski and Sobczyk}(2006)}]{Ankowski:2005wi}
\bibinfo{author}{\bibfnamefont{A.~M.} \bibnamefont{Ankowski}} \bibnamefont{and}
  \bibinfo{author}{\bibfnamefont{J.~T.} \bibnamefont{Sobczyk}},
  \bibinfo{journal}{Phys. Rev. C} \textbf{\bibinfo{volume}{74}},
  \bibinfo{pages}{054316} (\bibinfo{year}{2006}), \eprint{nucl-th/0512004}.

\bibitem[{\citenamefont{Ankowski and Sobczyk}(2008)}]{Ankowski:2007uy}
\bibinfo{author}{\bibfnamefont{A.~M.} \bibnamefont{Ankowski}} \bibnamefont{and}
  \bibinfo{author}{\bibfnamefont{J.~T.} \bibnamefont{Sobczyk}},
  \bibinfo{journal}{Phys. Rev. C} \textbf{\bibinfo{volume}{77}},
  \bibinfo{pages}{044311} (\bibinfo{year}{2008}), \eprint{0711.2031}.

\bibitem[{\citenamefont{Ankowski et~al.}(2012)\citenamefont{Ankowski, Benhar,
  Mori, Yamaguchi, and Sakuda}}]{Ankowski:2011ei}
\bibinfo{author}{\bibfnamefont{A.~M.} \bibnamefont{Ankowski}},
  \bibinfo{author}{\bibfnamefont{O.}~\bibnamefont{Benhar}},
  \bibinfo{author}{\bibfnamefont{T.}~\bibnamefont{Mori}},
  \bibinfo{author}{\bibfnamefont{R.}~\bibnamefont{Yamaguchi}},
  \bibnamefont{and} \bibinfo{author}{\bibfnamefont{M.}~\bibnamefont{Sakuda}},
  \bibinfo{journal}{Phys. Rev. Lett.} \textbf{\bibinfo{volume}{108}},
  \bibinfo{pages}{052505} (\bibinfo{year}{2012}), \eprint{1110.0679}.

\bibitem[{\citenamefont{Ankowski and Benhar}(2013)}]{Ankowski:2013gha}
\bibinfo{author}{\bibfnamefont{A.~M.} \bibnamefont{Ankowski}} \bibnamefont{and}
  \bibinfo{author}{\bibfnamefont{O.}~\bibnamefont{Benhar}},
  \bibinfo{journal}{Phys. Rev. D} \textbf{\bibinfo{volume}{88}},
  \bibinfo{pages}{093004} (\bibinfo{year}{2013}), \eprint{1305.2068}.

\bibitem[{\citenamefont{Benhar et~al.}(2008)\citenamefont{Benhar, day, and
  Sick}}]{Benhar:2006wy}
\bibinfo{author}{\bibfnamefont{O.}~\bibnamefont{Benhar}},
  \bibinfo{author}{\bibfnamefont{D.}~\bibnamefont{day}}, \bibnamefont{and}
  \bibinfo{author}{\bibfnamefont{I.}~\bibnamefont{Sick}},
  \bibinfo{journal}{Rev. Mod. Phys.} \textbf{\bibinfo{volume}{80}},
  \bibinfo{pages}{189} (\bibinfo{year}{2008}), \eprint{nucl-ex/0603029}.

\bibitem[{\citenamefont{McDonough}(2003)}]{MCDONOUGH2003547}
\bibinfo{author}{\bibfnamefont{W.}~\bibnamefont{McDonough}}, in
  \emph{\bibinfo{booktitle}{Treatise on Geochemistry}}, edited by
  \bibinfo{editor}{\bibfnamefont{H.~D.} \bibnamefont{Holland}}
  \bibnamefont{and} \bibinfo{editor}{\bibfnamefont{K.~K.}
  \bibnamefont{Turekian}} (\bibinfo{publisher}{Pergamon},
  \bibinfo{address}{Oxford}, \bibinfo{year}{2003}), pp.
  \bibinfo{pages}{547--568}, ISBN \bibinfo{isbn}{978-0-08-043751-4},
  \urlprefix\url{https://www.sciencedirect.com/science/article/pii/B0080437516020156}.

\bibitem[{\citenamefont{Lundberg and Edsjo}(2004)}]{Lundberg:2004dn}
\bibinfo{author}{\bibfnamefont{J.}~\bibnamefont{Lundberg}} \bibnamefont{and}
  \bibinfo{author}{\bibfnamefont{J.}~\bibnamefont{Edsjo}},
  \bibinfo{journal}{Phys. Rev. D} \textbf{\bibinfo{volume}{69}},
  \bibinfo{pages}{123505} (\bibinfo{year}{2004}), \eprint{astro-ph/0401113}.

\bibitem[{\citenamefont{Gluscevic and Boddy}(2018)}]{Gluscevic:2017ywp}
\bibinfo{author}{\bibfnamefont{V.}~\bibnamefont{Gluscevic}} \bibnamefont{and}
  \bibinfo{author}{\bibfnamefont{K.~K.} \bibnamefont{Boddy}},
  \bibinfo{journal}{Phys. Rev. Lett.} \textbf{\bibinfo{volume}{121}},
  \bibinfo{pages}{081301} (\bibinfo{year}{2018}), \eprint{1712.07133}.

\bibitem[{\citenamefont{Xu et~al.}(2018)\citenamefont{Xu, Dvorkin, and
  Chael}}]{Xu:2018efh}
\bibinfo{author}{\bibfnamefont{W.~L.} \bibnamefont{Xu}},
  \bibinfo{author}{\bibfnamefont{C.}~\bibnamefont{Dvorkin}}, \bibnamefont{and}
  \bibinfo{author}{\bibfnamefont{A.}~\bibnamefont{Chael}},
  \bibinfo{journal}{Phys. Rev. D} \textbf{\bibinfo{volume}{97}},
  \bibinfo{pages}{103530} (\bibinfo{year}{2018}), \eprint{1802.06788}.

\bibitem[{\citenamefont{Ooba et~al.}(2019)\citenamefont{Ooba, Tashiro, and
  Kadota}}]{Ooba:2019erm}
\bibinfo{author}{\bibfnamefont{J.}~\bibnamefont{Ooba}},
  \bibinfo{author}{\bibfnamefont{H.}~\bibnamefont{Tashiro}}, \bibnamefont{and}
  \bibinfo{author}{\bibfnamefont{K.}~\bibnamefont{Kadota}},
  \bibinfo{journal}{JCAP} \textbf{\bibinfo{volume}{09}}, \bibinfo{pages}{020}
  (\bibinfo{year}{2019}), \eprint{1902.00826}.

\end{thebibliography}

 \end{document}